\documentclass[pra,twocolumn,amsmath,amssymb,byrevtex]{revtex4}

\usepackage{bm,amssymb,amsmath,graphicx,color}

\newtheorem{Thm}{Theorem}\newtheorem{Prop}{Proposition}\newtheorem{Lem}{Lemma}\newtheorem{Cor}{Corollary}\newtheorem{Def}{Definition}\newcommand{\tr}{\mathop{\mathrm{tr}}\nolimits}\newcommand{\HA}{\mathop{\mathcal{H}}\nolimits}\newcommand{\D}{\mathop{\mathcal{D}}\nolimits}\newcommand{\LA}{\mathop{\mathcal{L}}\nolimits}\newcommand{\SA}{\mathop{\mathcal{S}}\nolimits}\newcommand{\M}{\mathop{\mathcal{M}}\nolimits}\newcommand{\R}{\mathop{\mathbb{R}}\nolimits}\newcommand{\I}{\mathop{\mathbb{I}}\nolimits}\newcommand{\A}{\mathop{\mathcal{A}}\nolimits}\newcommand{\E}{\mathop{\mathcal{E}}\nolimits}
\newcommand{\PA}{\mathop{\mathcal{P}}\nolimits}

\newcommand{\bra}[1]{\langle #1 |}\newcommand{\ket}[1]{| #1 \rangle}
\newcommand{\bracket}[2]{\langle #1 | #2 \rangle}\newcommand{\ketbra}[2]{| #1 \rangle \langle #2 |}
\begin{document}

\title{Distinguishability Measures and Entropies for General Probabilistic Theories}
\author{Gen Kimura ${}^{[a]}$}
\email{gen-kimura[at]aist.go.jp}
\author{Koji Nuida ${}^{[a]}$}
\email{k.nuida[at]aist.go.jp}
\author{Hideki Imai ${}^{[a],[b]}$}
\affiliation{[a] Research Center for Information Security (RCIS),
National Institute of Advanced Industrial
Science and Technology (AIST). 
Daibiru building 1003,
1-18-13 Sotokanda, Chiyoda-ku, Tokyo, 101-0021, Japan \\ 
[b] Graduate School of Science and Engineering,
Chuo University.
1-13-27 Kasuga, Bunkyo-ku, Tokyo 112-8551, Japan
}

\begin{abstract}
As a part of the construction of an information theory based on general probabilistic theories, we propose and investigate the several distinguishability measures and ``entropies" in general probabilistic theories. 
As their applications, no-cloning theorems, information-disturbance theorems are reformulated, and a bound of the accessible informations is discussed in any general probabilistic theories, not resorting to quantum theory. 
We also propose the principle of equality for pure states which makes general probabilistic theories to be more realistic, and discuss the role of entropies as a measure of pureness. 
\end{abstract}
\maketitle

\section{Introduction}

Recent development of the quantum information theory has shown us the ability 
of information processings and computations based on the quantum physics can go far beyond 
those based on classical physics. 
At its heart, this is because the potential ability of a probability is enlarged from classical theory to quantum theory. 
Indeed, quantum theory can be considered as a probabilistic theory, which --- in some sense --- properly includes the classical probability theory (Kolomogorov's probability theory). 
However, this does not mean that quantum theory is the most general theory of a probability even among the possible theories which have an operational meanings. 
So far, the most general theory of a probability with a suitable operational meanings has been developed by several researchers (See for instance \cite{ref:Mackey,ref:Gudder1,ref:HolevoSM,ref:Ludwig,ref:Davies,ref:Ozawa,ref:Barrett}). 
Following the recent trend, we call such theories the {\it general probabilistic theories} (or simply GPTs). 

As the quantum information theory has been constructed based on the quantum theory, information theories can be constructed based on each probabilistic theory \cite{ref:Barrett2,ref:Gisin,ref:Barrett,ref:Dariano,ref:KMI,ref:ZP,ref:NKM,ref:Barnum}. 
There are several motivations for this line of researches: 
First, this is an attempt to find physical principles (axioms written by physical languages) for quantum theory \cite{ref:Jammer,ref:QL,ref:algebra}. 
Indeed, by considering the general framework which encompasses the quantum theory, we look for principles which determine the position of the quantum theory in this general framework. 
The development of the quantum information theory motivate us to find the principles based on information processings for the theory of quantum physics \cite{ref:Fuchs02,ref:Clifton,ref:Dariano}. 
Second, the construction of the information theory based on the most general theory of probability enables us to understand logical connections among information processings by resorting to the particular properties of neither classical nor quantum theory, but only to the essential properties which a suitable probability theory should possess. 
Third, this is a preparation for the possible break of quantum theory. 
For instance, one can discuss a secure key distribution in the general framework without assuming quantum theory itself \cite{ref:KD}.   
Finally, this might provide a classical information theory under some restrictions of measurements, since any general probabilistic theories has a classical interpretation based on such restrictions of measurements \cite{ref:HolevoSM,ref:Ginp}.  

In this paper, we propose and give systematic discussions of several distinguishability measures (especially, {\it Kolmogorov distance} and {\it fidelity}) and three quantities related to entropies for general probabilistic theories.  
The corresponding measures and entropies in classical and quantum theory have been proved to be useful \cite{ref:FG,ref:NC}, and we give generalizations for them in any GPTs and discuss their applications. 
In particular, no-cloning theorem and a simple information-disturbance theorem in GPTs are reformulated using fidelity, and a bound of the accessible information is discussed based on one of the ``entropies". 
Finally, we introduce and formulate the principle of ``equality of pure states" meaning that there are no special pure states. 
We call such GPT symmetric and in symmetric GPT, the measure of pureness will be discussed. 

\section{General Probabilistic Theories}\label{sec:GPT}
 
In this section, we give a brief review of general probabilistic theories (See for instance \cite{ref:Mackey,ref:Gudder1,ref:HolevoSM,ref:Barrett} and references therein for details.)  
Although, in the end, we are going to use mathematical notions such as convexity, affine functions, etc., it should be noticed that we do not assume any mathematical structure without physical reasons. 

The important ingredients of the GPTs are the notions of state and measurement. 
In any GPT, we have a physical law to determine a probability $p(a | M,s)$ to obtain an output $a$ by a measurement $M$ of an observable under a state $s$. In this paper, for simplicity, we only treat a measurement with a finitely  many outcomes. 
Naturally, we assume the separating properties of both states and measurements: (A1) States $s_1$ and $s_2$ are identified if $p(a|M,s_1) = p(a|M,s_2)$ for any measurement $M$ and measurement outcome $a$; 
(A2) Measurement $M_1$ and $M_2$ are identified if $p(a|M_1,s) = p(a|M_2,s)$ for any measurement outcome $a$ under any state $s$. 
We also assume the convex property of states; 
(A3) For any states $s_1,s_2$ and $q \in [0,1]$, there exists the state $s$ to prepare $s_1$ with probability $q$ and $s_2$ with probability $1-q$; namely, it follows that $p(a | M,s) = q p(a| M,s_1) + (1-q) p(a| M,s_2)$ for any measurement; 
(A4) Further, we naturally assume that the dynamics preserves this probabilistic mixtures; 
(A5) We introduce a natural topology on the state space which is the weakest topology such that $s \to p(a | M,s) $ is continuous for any measurements; Finally, we assume (A6) a joint state $\omega$ of system $A$ + $B$ defines a joint probability for each measurements $M_A$ and $M_B$ which satisfies the no-signaling condition, i.e., the marginal probabilities for the outcomes of a measurement on $A$ do not depend on the measurement choices on $B$, and vice versa. 
Moreover, the joint state is determined by joint probabilities for all pairs of measurements of $A$ and $B$. 

Based on these, one can show the followings \cite{ref:Mackey,ref:Gudder1,ref:HolevoSM,ref:Barrett}: 

(a) There exists a locally convex topological vector space $V$ such that, in a suitable representation, {\it the state space $\SA$ is a convex subset in $V$ where $q s_1 + (1-q) s_2$ corresponds to the state described in (A3) above}. 
An extreme point of $\SA$ is called a pure state.  
Moreover, without loss of generality, one can assume that $\SA$ is compact with a natural topology \cite{ref:NKM}. 
Notice that by the famous Krein-Milman theorem (see, for instance, Theorem 10.4 in \cite{ref:Schaefer}) the set of extreme points $\SA_{pure}$ is non-empty and $\SA$ is the closed convex hull of extreme points. 
In particular, in finite dimensional cases, any state $s \in \SA$ has a convex decomposition with finite numbers of pure states (hereafter, {\it a pure state decomposition}): $s = \sum_x p_x s_x$ where $p_x \ge 0, \sum_x p_x = 1, s_x \in \SA_{pure}$ (see, for instance, Theorem 5.6 in \cite{ref:Lay}). 
 
A map $f : \SA \to \R$ is called an affine functional if it satisfies $f(q s_1 + (1-q)s_2) = q f(s_1) + (1-q)f(s_2)$ for any $q \in [0,1], s_1,s_2 \in \SA$. 
In particular an affine functional $e: \SA \to \R$ is called an effect if the range is contained in $[0,1]$. 
We denote the sets of all the affine functional and all the effects by $\A(\SA)$ and $\E(\SA)$, respectively. 
It is easy to see that $\E(\SA)$ is a convex subset of a real vector space $\A(\SA)$. 
We call an extreme effect {\it a pure effect}.  
The zero effect $0$ and unit effect $u$ such that $0(s) = 0$ and $u(s) = 1$ are trivially pure effects. 
It is easy to see that effect $u -e $ is pure iff effect $e$ is pure. 
Moreover, we can introduce a natural topology on $\E(\SA)$ which is the weakest topology such that the map $\E(\SA) \to \R$, $e \mapsto e(s)$, becomes continuous for every $s \in \SA$.
One can that $\E(\SA)$ is compact with respect to this topology \cite{ref:NKM}.

(b) It is often convenient to characterize a measurement without explicitly specifying the measurement outcomes. 
In that case, {\it any measurement $M$ is characterized by the set of effects $m_i$ such that $p(a_i | M,s) = m_i(s)$ and $\sum_i m_i = u $}: 
In the following, we occasionally use the notation $M = (m_j)_j$ (implicitly assuming conditions $m_j \in \E(\SA)$ and $\sum_j m_j = u$) to denote the measurement on $\SA$ meaning that $m_j(s)$ is the probability to obtain $j$th output (say $a_j$) by a measurement $M$ under a state $s$. 

(c) {\it Dynamics is described by an affine function on state space. }
In general, the initial state space $\SA$ and final state space $\SA^\prime$ might be different. Then, a time evolution map is given by an affine map $f$ from $\SA$ to $\SA^\prime$. We denote by $\A(\SA,\SA^\prime)$ the set of all the affine map from $\SA$ to $\SA^\prime$. 

(d) The joint systems are described by a convex set in a tensor product of the corresponding vector spaces. 
A joint state $\omega$ on $A+B$ with state spaces $\SA_A$ and $\SA_B$ is described by a bi-affine map on $\E(\SA_A) \times \E(\SA_B)$. 
In particular, if $\omega$ is a joint state on $A+B$, then the marginal state of $A$ is defined by $\omega_A(e) := \omega(e, u_B) \ (e \in \E(\SA_A))$ where $u_B$ is the unit effect on $\SA_B$. 
From the extreme property of pure states, it is easy to see \footnote{Although the proof is simple (see for instance \cite{ref:Take,ref:Barrett}), this property is important in its applications. 
For instance, in the context of key distribution, Alice and Bob can assure to be safe if there joint state is pure, since then their system does not have any correlations with another system (eavesdropper). } that if 
the marginal state $\omega_A $ is pure, then a joint state $\omega$ is a state with no correlations: $\omega(e_A,e_B) = \omega_A(e_A) \omega_B(e_B) \ (e_A\in \E(\SA_A),e_B\in \E(\SA_B))$. 
 
It is important to notice that all the mathematical structures are not introduced {\it ad hoc} but they appear naturally based on physical assumptions (A1-A6). 
It is also possible to formulate the measurement process by considering the cone generated by $0$ and $\SA$ in $V$ \cite{ref:Gudder1,ref:Dariano,ref:Davies}. 

In this paper, we treat for simplicity finite GPT where $V$ is finite dimensional, but most of the definitions and properties below holds with some topological remarks \cite{ref:Schaefer}. 
(However, notice that in finite dimensional cases, there are essentially the unique topology, and one can use another characterization of the natural topology, for instance using the Kolmogorov distance below. 
In particular, the unique topology is the Euclidean topology and thus one can imagine a state space of each GPT as any compact convex (or equivalently closed bounded convex) subset in Euclidean spaces.) 
Moreover, we assume that any set of effects $m_i$ such that $\sum_i m_i = u$ has a correspondent measurement. 
(It is also easy exercises to reformulate below without this assumption.) 

Here, let us see the typical examples for finite GPTs. 

[Finite Classical Systems] Let 
$\Omega = \{\omega_1,\ldots,\omega_d\}$ be a sample space. 
The state is represented by a probability $p_i$ for an elementary event $\omega_i$. 
The state space is given by $\SA_c := \{{\bm p} \in \R^d | p_i \ge 0, \sum_{i=1}^d p_i=1\}$. 
There are $d$ numbers of pure states, which are the definite states where one of the elementary event occurs with probability $1$: Namely, 
${\bm p^{(\mu)}}=(\delta_{\mu 1},\ldots,\delta_{\mu d}) \in \SA_c \ (\mu=1,\ldots,d)$D
Notice that $\SA_c$ is a (standard) simplex. 
In particular, any state ${\bm p} = (p_1,\ldots,p_d) \in \SA_c$ has the unique pure state decomposition: ${\bm p} = \sum_{\mu =1}^d p_\mu  {\bm p^{(\mu)}}$.

[Finite Quantum Systems] Let $\HA$ be a $d$ dimensional complex Hilbert space. 
A quantum state is represented by a density operator $\rho$, i.e., a positive operator on $\HA$ with unit trace. 
The state space is given by $\SA_q = \{ \rho \in \LA(\HA) \ | \ \rho \ge 0, \tr \rho= 1\}$ where $\LA(\HA)$ is a real vector space of all the (Hermitian) operator on $\HA$. 
Pure states are characterized by $1$ dimensional projection operators. 
A quantum effect $e$ is represented by an operator $E$ satisfying $0 \le E \le \I$, called a POVM (positive operator valued measure) element, by the correspondence $e(\rho) = \tr (\rho E)$. 
Here $0,\I$ denote the zero and identity operator on $\HA$. 
In particular, any measurement of an observable $M = (m_i)_i $ where $m_i$ are effects on $\SA_q$ has the correspondent POVM measurement $(M_i)_i$ such that $M_i \in \LA(\HA), \ M_i \ge 0, \ \sum_i M_i = \I$ and $m_i(\rho) = \tr (\rho M_i)$.  
Notice that the set of all the extreme effects is the set of all the projection operator $\PA(\HA)$. 
The POVM measurement $(P_i)_i$ consists of projection operators $P_i$ is called a PVM (projection valued measure) measurement.
The following is an example of GPT which is neither classical nor quantum: 

[Hyper Cuboid Systems and squared system] Let $\SA_{cb} := \{{\bm c} \in \R^d | 0 \le c_i \le 1 (i=1,\ldots,d)\}$. 
The pure states are $2^d$ numbers of vertexes. 
We call this {\it hyper cuboid system} and especially the {\it squared system} when $d = 2$ \cite{ref:KMI}. 
These might be the easiest examples of GPT which are neither classical nor quantum. 
However, one can construct a classical model such that a suitable restriction of measurements reduces the hyper cuboid systems \cite{ref:Ginp}.  
 
Finally, notice that the probabilistic theories with state spaces $\SA_A$ and $\SA_B$ are equivalent if they are affine isomorphic, i.e., there exists a bijective affine map from $\SA_A$ to $\SA_B$. 
For instance, any GPT which has a simplex state space is affine isomorphic to some standard simplex, and therefore can be considered as a classical system.

\section{Distinguishability Measures for General Probabilistic Theories}\label{sec:DM}

In this section, we introduce several distinguishability measures (Kolmogorov distance, Fidelity, Shannon distinguishability etc) for GPTs. 
The corresponding measures for quantum systems are proved to be useful in quantum information theories \cite{ref:FG}.  
It is indeed straightforward to generalize them to any GPT using the notions developed in the preceding sections, and some of them has been used in references \cite{ref:Dariano,ref:ZP,ref:NKM}. 
Most of the properties for quantum systems preserves to be hold including the ways to prove them \cite{ref:FG}. 
However, we think it useful to sum up these measures, especially Kolmogorov distance and fidelity, for GPT systematically and all the proofs of this section are put in Appendix \ref{app:proofs} for the reader's convenience. 
A striking thing is that all the below results does not resort to ingredients such as vectors and operators on a Hilbert space, but only to the analysis of probabilities. 

All the measures below are based on those for classical systems among every possible measurements of observables: In the following, let $\SA$ be the set of states (state space), and $\E= \E(\SA)$, $\M = \M(\SA)$ be the sets of effects and measurements on $\SA$. 

\subsection{Kolmogorov Distance in GPT}

The Kolmogorov distance $D_c({\bm p},{\bm q})$ is known to serve as a good distinguishability measure between two probability distributions ${\bm p}= (p_i)_i$ and ${\bm q} = (q_j)_j$ : 
$$
D_c({\bm p},{\bm q}) := \frac{1}{2} \sum_i |p_i - q_i|. 
$$
Indeed, $D_c$ has a metric property and it follows that $D_c(p_i,q_j) = \max_S |p(S) - q(S)|$ where the maximization is taken over all subsets $S$ of the index set $\{i\}$. 
Thus $D_c({\bm p},{\bm q})$ is considered as a metric for two probability distributions with an operational meaning.  

In any GPT, one can define \cite{ref:NKM} the Kolmogorov distance between two states $s_1,s_2 \in \SA$ by 
\begin{equation}\label{eq:D}
D(s_1,s_2) := \max_{M=(m_i) \in \M} D_c({\bm p}_1(M) ,{\bm p}_2(M)), 
\end{equation}
where ${\bm p}_1(M) := (m_i(s_1))_i $ and ${\bm p}_2(M) := (m_i(s_2))_i $ are probability distributions to get $i$th output of the measurement $M$ under states $s_1$ and $s_2$, respectively. 
The maximization in \eqref{eq:D} is always attained by some measurement, which we call an optimal measurement, due to the compactness of the effect set \cite{ref:NKM}. 
Notice that $D(s_1,s_2)$ is a metric of ${\cal S}$, i.e., (i) $D(s_1,s_2) \ge 0 \ $; equality iff $s_1 = s_2$, (ii)$D(s_1,s_2) =  D(s_2,s_1)$, and (iii) $D(s_1,s_3) \le D(s_1,s_2) + D(s_2,s_3)$, and it is bounded above from $1$, i.e., $0 \le D(s_1,s_2) \le 1$. 
These follow from a metric property of $D_c$ and a separation property of states. 
The above mentioned operational meaning of $D_c$ also gives $D(\rho,\sigma) $ an operational meaning; that is the maximum difference of probability among all the event $S$ and all measurements.  
In quantum systems, $D(\rho_1,\rho_2)$ is the trace distance between density operators $\rho_1,\rho_2$: $D(\rho_1,\rho_2) = \frac{1}{2}\tr |\rho_1 - \rho_2|$ \cite{ref:NC} where $|A|:= \sqrt{A^\dagger A}$.  

For any measurement $M=(m_i)$ and states $s_1,s_2 \in \SA$, one can consider a two valued measurement $M_2 = (m_+,m_-)$ where $m_+ := \sum_{i \in X_+} m_i$ and $m_- := \sum_{i \in X_-} m_i$ with $X_+ := \{ i \ | \ m_i(s_1) - m_i (s_2) \ge 0\}$ and $X_- :=\{ i \ | \ m_i(s_1) - m_i (s_2) < 0\}$. 
Using this, one has another characterization of the Kolmogorov distance: 
\begin{equation}\label{eq:DA}
D(\rho,\sigma) = \max_{e \in \E} (e(\rho) - e(\sigma) ). 
\end{equation}
The quantity in the right-hand side is a metric used in \cite{ref:Dariano}. 

Let $P_s(s_1,s_2)$ be the maximal success probability to distinguish two states $s_1$ and $s_2$ in a single measurement under the uniform prior distribution. 
Without loss of generality, it is enough to consider two-valued measurement $(m_1,m_2) \in \M$ for a discrimination problem of two states $s_1$ and $s_2$ by guessing $s_1$ (or $s_2$) when observing $1$ (or $2$)th output. 
Thus, we have   
\begin{eqnarray}\label{eq:Ps}
P_s(s_1,s_2) &:=& \max_{(m_1,m_2) \in \M} \Bigl(\frac{1}{2} m_1(s_1)  + \frac{1}{2} m_2(s_2) \Bigr) \nonumber \\
&=& \frac{1}{2} \Bigl( 1 + \max_{e\in \E} (e(s_1) - e(s_2))\Bigr). 
\end{eqnarray}
From \eqref{eq:DA} and \eqref{eq:Ps}, we have another operational meaning of the Kolmogorov distance: 
\begin{Prop}\label{prop:OMofD} For any states $s_1,s_2 \in \SA$ in GPT, 
$$
D(s_1,s_2) = 2 P_s(s_1,s_2) - 1. 
$$
\end{Prop}
Note that $D(s_1,s_2)$ takes the maximum $1$ iff $P_s(s_1,s_2) = 1$, i.e., when 
$s_1$ and $s_2$ are completely distinguishable in a single measurement. 
On the other hand, $D(s_1,s_2)$ takes the minimum $0$ (thus $s_1=  s_2$) iff $P_s(s_1,s_2) = 1/2$, i.e., $s_1$ and $s_2$ are completely indistinguishable (and indeed such states should be identified due to the separation property of states). 

In the following, we show the monotonicity, strong convexity, joint convexity, and convexity follow for the Kolmogorov distance in any GPT. 
\begin{Prop}\label{prop:monD} (Monotonicity) For any states $s_1,s_2 \in {\cal S}$, and time evolution map $\Lambda \in \A(\SA,\SA^\prime)$, we have  
$$
D(s_1,s_2) \ge D(\Lambda(s_1),\Lambda(s_2)). 
$$
\end{Prop}
This implies that the distinguishability between $s_1$ and $s_2$ cannot be increased in any physical means.  
Notice that it is well known that the trace distance in quantum systems has the monotonicity property under any trace preserving completely positive map \cite{ref:NC}. 
Proposition \ref{prop:monD} generalizes this for any trace preserving positive map. 

\begin{Prop}\label{prop:sc}(Strong convexity) Let ${\bm p} = (p_i)_i$ and ${\bm q}=(q_i)_i$ be probability distributions over the same index set, and 
$s_i, t_i \in \SA$ be states of GPT with the same index set. Then, it follows that 
$$
D\Bigl( \sum_i p_i s_i, \sum_i q_i t_i \Bigr) \le D_c({\bm p},{\bm q}) + \sum_i p_i D(s_i,t_i). 
$$
\end{Prop}
As corollaries, we have 
\begin{Cor}(Joint convexity) 
$$
D\Bigl( \sum_i p_i s_i, \sum_i p_i t_i \Bigr) \le \sum_i p_i D(s_i,t_i). 
$$
(As a special case $p_i = q_i$ of the strong convexity.)
\end{Cor}
\begin{Cor}
(Convexity) 
$$
D\Bigl( \sum_i p_i s_i, t \Bigr) \le \sum_i p_i D(s_i,t). 
$$
(As a special case $t_i = t$ of the joint convexity.)
\end{Cor} 

\subsection{Fidelity in GPT} 

The Bhattacharyya coefficient (the classical fidelity) between two probability distributions ${\bm p}= (p_i)_i$ and ${\bm q} = (q_j)_j$ is defined by :
\begin{equation}
F_c({\bm p},{\bm q}) := \sum_i \sqrt{p_i q_i}.  
\end{equation}
Note that (i) $0 \le F_c({\bm p},{\bm q}) \le 1$ where $F_c({\bm p},{\bm q}) = 1 $ iff ${\bm p}={\bm q}$; (ii) $F_c({\bm p},{\bm q}) = F_c({\bm q},{\bm p})$. 
We say two probability distributions ${\bm p},{\bm q}$ are orthogonal iff $F_c({\bm p},{\bm q}) = 0$. 

In any GPT, one can also define the fidelity \cite{ref:Gudder1,ref:ZP} between two states $s_1,s_2 \in \SA$ as  
\begin{equation}\label{eq:Fidelity}
F(s_1,s_2) = \inf_{M=\{m_i\} \in \M } F_c({\bm p}_1(M) ,{\bm p}_2(M)),  
\end{equation}
where ${\bm p}_1(M) := (m_i(s_1))_i $ and ${\bm p}_2(M) := (m_i(s_2))_i $. 
Contrast to the Kolmogorov distance, the attainability of the infimum of the fidelity seems to be nontrivial.  
In quantum mechanics, one has the formula $F(\rho_1,\rho_2) = \tr |\rho_1^{1/2}\rho_2^{1/2}| = \tr[(\rho_1^{1/2}\rho_2 \rho_1^{1/2})^{1/2}]$ 
between two density operators $\rho_1,\rho_2$ \cite{ref:NC,ref:FC}. 
Also, it is shown that an optimal measurement (POVM) exists which attains the infimum.  

From the property of the Bhattacharyya coefficient and the separation property of states, 
it follows that (i) $0 \le F(s_1,s_2) \le 1$ where
$ F(s_1,s_2) = 1$ iff $s_1 = s_2$; (ii) $F(s_1,s_2) = F(s_2,s_1)$. 
We say that states $s_1$ and $s_2$ are orthogonal ($s_1 \perp s_2$) iff $F(s_1,s_2) = 0$. 
\begin{Prop}\label{prop:monF} (Monotonicity) 
For any states $s_1,s_2 \in {\cal S}$, and time evolution map $\Lambda \in \A({\cal S},{\cal S}^\prime)$, it follows 
$$
F(\Lambda(s_1),\Lambda(s_2)) \ge F(s_1,s_2). 
$$
\end{Prop}

\begin{Prop}\label{prop:scF}(Strong concavity \cite{ref:ZP}) Let ${\bm p}=(p_i)_i$ and ${\bm q} = (q_i)_i$ be probability distributions over the same index set, and $s_i,t_i \in \SA$ be states of GPT with the same index set. Then, 
$$
F\Bigl( \sum_i p_i s_i, \sum_i q_i t_i \Bigr) \ge \sum_i \sqrt{p_iq_i} F(s_i,t_i). 
$$
\end{Prop}
As corollaries, one gets 
\begin{Cor}(Joint concavity and concavity)
\begin{eqnarray*}
F\Bigl( \sum_i p_i s_i, \sum_i p_i t_i \Bigr) \ge \sum_i p_i F(s_i,t_i), \\
F\Bigl( \sum_i p_i s_i, t \Bigr) \ge \sum_i p_i F(s_i,t). 
\end{eqnarray*}
\end{Cor}

\begin{Prop}\label{prop:biF} In a bipartite system $A + B$, we have the followings: 

(i) $F(s_A ,t_A) \ge F(s,t)$ for any $s,t \in {\cal S}_A\otimes {\cal S}_B$ where $s_A$ and $t_A$ are the reduced states to the system $A$. 

(ii) $ F(s_1,s_2) F(t_1 ,t_2) \ge F(s_1\otimes t_1 ,s_2\otimes t_2) $ for any $s_1,s_2 \in {\cal S}_A, t_1,t_2 \in {\cal S}_B$. 

(iii) $F(s_1,s_2) = F(s_1\otimes t,s_2 \otimes t) $ for any $s_1,s_2 \in {\cal S}_A, t \in {\cal S}_B$. 
\end{Prop}
In particular, from (ii), it follows 
\begin{equation}\label{eq:F2F}
F(s,t)^2 \ge F(s\otimes s,t \otimes t), 
\end{equation}
by letting $\SA:= \SA_A = \SA_B$ and $s=s_1=t_1 \in \SA$ and $t=s_2=t_2 \in \SA$. 

Note that the generalization of properties of Proposition \ref{prop:biF} is straightforward for multipartite system.  

However, contrary to the Kolmogorov distance, it is difficult to give an operational meaning for the Fidelity, since there is no known operational meaning of Bhattacharyya coefficient. 
In using the Fidelity, it is important to know the relation with another operational measures like the Kolmogorov distance.  

\subsection{Relation between the Kolmogorov Distance and the Fidelity} 

\begin{Prop}\label{prop:relGF} 
For any state $s, t \in \SA$, it follows 
\begin{equation}\label{eq:relGF}
1 - F(s,t) \le D(s,t) \le \sqrt{1 - F(s,t)^2}. 
\end{equation} 
\end{Prop}
This relation is famous to hold in quantum systems \cite{ref:FG,ref:NC}, but  
Proposition \ref{prop:relGF} shows that this holds for any GPT. 

From \eqref{eq:relGF}, we have 
\begin{Cor}\label{cor:EqFD}
(i) $D(s,t) = 0 $ iff $F(s,t) = 1$ and (ii) $D(s,t) = 1$ iff $F(s,t) = 0$. 
In particular, the orthogonality of states turns out to be equivalent to the complete distinguishability of states ($P_s = 1$). 
\end{Cor}
In this sense, the Kolmogorov distance and the fidelity is equivalent. 

Similarly, it is straightforward to introduce another measures which are used in quantum information theory. 
For instance, one can define Shannon distinguishability and can show the same relations (see for instance Theorem 1 in \cite{ref:FG}). 

\section{Applications} 

In this section, we give simple proofs using the Fidelity for no-cloning theorem \cite{ref:Barrett} and information-disturbance theorem \cite{ref:NKM,ref:ZP} in any GPT. 

\begin{Thm}\label{thm:NoClone1}  (No-cloning) 
In any GPT, two states $s_1,s_2 \in \SA$ are jointly clonable iff $s_1 = s_2$ or $s_1$ and $s_2 $ are completely distinguishable.  
\end{Thm}
{\bf Proof} Let states $s_1,s_2 \in \SA$ are jointly clonable. 
Namely, there exists a time evolution map (a cloning machine) $\Lambda \in \A(\SA,\SA\otimes \SA) $ satisfying 
\begin{equation}\label{eq:cl}
\Lambda(s_1) = s_1\otimes s_1, \ \Lambda(s_2) = s_2\otimes s_2. 
\end{equation}
From \eqref{eq:F2F}, we have 
$$
F(\Lambda(s_1),\Lambda(s_2)) = F(s_1 \otimes s_1,s_2 \otimes s_2) \le F(s_1,s_2)^2.  
$$ 
From the monotonicity of $F$, it follows that $F(s_1,s_2) \le F(\Lambda(s_1),\Lambda(s_2)) \le F(s_1,s_2)^2$, which implies that $F(s_1,s_2) = 0$ or $1$. 
In other words, $s_1 = s_2$ or $s_1$ and $s_2$ are completely distinguishable (cf. Corollary \ref{cor:EqFD}). 

Suppose that $s_1 = s_2 $, then one has a time evolution map $\Lambda \in \A(\SA,\SA\otimes \SA)$ defined by $\Lambda(s):= s \otimes s_1$. (Physically, this is nothing but a preparation of a fixed state $s_1$.)  
It is obvious that this jointly clones $s_1 $ and $s_2$.  
Next, suppose that $s_1 $ and $s_2$ are completely distinguishable. 
Namely, there exists a measurement $M=(m_1,m_2) \in \M(\SA)$ such that 
$m_1(s_1) = 1, m_1(s_2) = 0$ (and thus $m_2(s_1) = 0, m_2(s_2) = 1$). 
Then, $\Lambda (s) := m_1(s) s_1 \otimes s_1 + m_2(s) s_2 \otimes s_2$ for any $s \in \SA$ defines a time evolution map $\Lambda \in \A(\SA,\SA\otimes\SA)$ satisfying the cloning condition \eqref{eq:cl}. 
 (Notice that $m_1(s),m_2(s) \ge 0, m_1(s) + m_2(s) = 1$ and thus $m_1(s) s_1 \otimes s_1 + m_2(s) s_2 \otimes s_2 \in \SA \otimes \SA$ from the convexity of $\SA \otimes \SA$. The affinity of $\Lambda$ follows from the affinity of $m$.)  \hfill $\blacksquare$

\begin{Lem}\label{lem:notdis}
For any GPT with at least two distinct states, there exists two distinct states which are not completely distinguishable. 
\end{Lem}
{\bf Proof} \ Let $s_1 \neq s_2 \in \SA$. 
Assume that any two distinct states are completely distinguishable. 
Then, we have $F(s_1,s_2) = 0$. 
From the convexity of $\SA$, there exists a state $s:= \frac{1}{2} s_1 + \frac{1}{2} s_2 \neq s_1$. 
From the concavity of $F$, we have 
$F(s_1,s) \ge \frac{1}{2} F(s_1,s_1) + \frac{1}{2} F(s_1,s_2) = \frac{1}{2}$. 
Therefore, $s_1$ and $s$ are distinct states which are not completely distinguishable. 
\hfill $\blacksquare$

We call a physical process which clones any unknown states {\it a universal cloning machine} : 
\begin{Prop} (No-cloning) In any GPT with at least two distinct states, there are no universal cloning machine. 
\end{Prop}
{\bf Proof} \ This follows from Theorem \ref{thm:NoClone1} and Lemma \ref{lem:notdis}.  
\hfill $\blacksquare$

In a usual application, cloning is often considered for only pure states. 
We call a physical process which clones any unknown pure states {\it a universal cloning machine for pure states} : However, such cloning 
is possible if and only if GPT is classical: 
\begin{Prop} 
GPT is classical iff there is a universal cloning machine for pure states. 
\end{Prop}
{\bf Proof} \ Notice that classical systems are characterized by the fact that all the pure states are completely distinguishable \cite{ref:Barrett}. This fact and Theorem \ref{thm:NoClone1} complete the proof. \hfill $\blacksquare$

\begin{Thm} \label{thm:ID}(Information disturbance)
In any GPT, any attempt to get information to discriminate two pure states which are not completely distinguishable inevitably causes disturbance. 
\end{Thm}
{\bf Proof} Let $s_1,s_2 \in \SA_A$ be two pure states which are not completely distinguishable, i.e., $0 < F(s_1,s_2) $. 
Assume that there is a physical mean to get information to discriminate $s_1,s_2$ without causing any disturbance to the system. 
This implies that we have a time evolution map $\Lambda \in \A(\SA_A,\SA_A\otimes \SA_B)$ and initial states $t_0 \in \SA_B$ such that the reduced states to system A is the same:  
$$
\Lambda(s_1\otimes t_0)_A = s_1, \ \Lambda(s_2\otimes t_0)_A = s_2.  
$$
Since $s_1,s_2$ are pure states, there exists no correlations between system A an B, and hence one gets 
$$
\Lambda(s_1\otimes t_0) = s_1 \otimes t_1, \ \Lambda(s_2\otimes t_0) = s_2 \otimes t_2, 
$$
for some $t_1,t_2 \in \SA_B$. 
From the monotonicity of $F$ and Proposition \ref{prop:biF}, it follows that 
$F(s_1,s_2) = F(s_1\otimes t_0,s_2\otimes t_0) \le F(\Lambda(s_1\otimes t_0),\Lambda(s_2\otimes t_0)) = F(s_1,s_2) F(t_1,t_2)$. Since $0 < F(s_1,s_2)$, we have $F(t_1,t_2) = 1$ and thus $t_1 = t_2$. 
Therefore, to get information to distinguish $s_1 $ and $s_2$, one has to inevitably disturb at least one of these states.  \hfill $\blacksquare$ 

No cloning theorems are discussed in \cite{ref:Barrett} with completely different methods. In \cite{ref:NKM}, we have proved Theorem \ref{thm:ID} using the Kolmogorov distance. 
Essentially the same proof as above is given in \cite{ref:ZP}. 

\section{Indecomposable and Complete measurement in General Probabilistic Theories}

\subsection{Indecomposable Effect}

In quantum systems, a fundamental POVM element $E$ is that with one dimensional range, called a rank-one POVM element.  
Let us define the corresponding notions in any GPT, which we are going to call an {\it indecomposable} effect:  
\begin{Def}\label{def:indec}
We call an effect $e \in \E(\SA)$ {\it indecomposable} if (i) $e \neq 0$ and (ii) for any decomposition $e = e_1 + e_2$ into the sum of two effects $e_1$ and $e_2$, there exists $c \in \R$ such that $e_1 = c e$. 
We denote the set of all the indecomposable effects on $\SA$ by $\E_{ind}(\SA) \subset \E(\SA)$.  
\end{Def} 
It is easy to see that the above mentioned $c$ satisfies $0 \le c \le 1$. 

Here, we show some general properties of effects and indecomposable effects: 
\begin{Prop}\label{prop:PEzeros}
Let $e$ be a non-zero pure effect on $\SA$. 
Then, there exists a state $s$ such that $e(s) = 1$. 
Since $\SA$ is compact, such state can be taken to be a pure state. 
\end{Prop}
{\bf Proof} \ 
Suppose that there are no state $s$ such that $e(s) = 1$. 
Then, from the compactness of $\SA$, we have   
$$
\sup_{s \in \SA} e(s) = \max_{s\in \SA} e(s) =: x<  1. 
$$
From this, $\tilde{e} := e/x$ is an effect which is neither $e$ nor zero effect $0$. 
Since we have the identity, 
$$
e = x \tilde{e} + (1-x) 0,
$$
this contradicts that $e$ is a pure effect. 

Let $s = \sum_i p_i s_i \ (p_i > 0, \sum_i p_i = 1)$ be a pure state decomposition of $s$. 
Then, it is easy to see $e(s_i) = 1$ for any pure state $s_i$. 
Thus, we can take a pure state $s$ such that $e(s) = 1$. \hfill $\blacksquare$

\begin{Cor}
Let $e$ be a pure effect which is not $u$. 
Then, there exists a state $s$ such that $e(s) = 0$. Such state can be taken to be a pure state. 
\end{Cor}
{\bf Proof} \ 
Since $e (\neq u)$ is pure, the effect $\tilde{e} = u-e$ is non-zero pure effect. 
From Proposition \ref{prop:PEzeros}, there exits a pure state $s$ such that $\tilde{e}(s) = 1 - e(s) = 1$. 
Thus, $e(s) = 0$. \hfill $\blacksquare$

Next, we show that any non-zero effect has a decomposition with respect to indecomposable effects:  
\begin{Prop}\label{lem:EdecInd}
In any GPT, for every $ 0 \neq e  \in \E(\SA)$, there exist a finite collection of indecomposable effects $e_i \in \E(\SA), 1 \le i \le r$, such that $e = \sum_{i=1}^r e_i$. 
In particular, in any GPT, there exists an indecomposable effect. 
\end{Prop} 
(See Appendix A for the proof.)
Moreover, we have: 
\begin{Prop}\label{prop:InPureEff}
In any GPT, there exists an indecomposable and pure effect. 
\end{Prop}
{\bf Proof} \ 
To prove this, we use the following lemmas: 

\begin{Lem}\label{lem:ietoie}
Let $e \in \E$ be an indecomposable effect and let $q:= \max_{s \in \SA} e(s) $. (Note that $0 < q \le 1$.) 
Then, $\tilde{e} := \frac{1}{q}$ is an indecomposable effect.   
\end{Lem}
\begin{Lem}\label{lem:inep}
If $e$ is indecomposable effect such that there exists a state $s$ satisfying $e(s) = 1$, then $e$ is a pure effect.  
\end{Lem}
(See appendix A for the proofs.) 
From Proposition \ref{lem:EdecInd}, there exists an indecomposable effect. 
From Lemma \ref{lem:ietoie}, one can construct an indecomposable effect from any indecomposable effect such that $\tilde{e}(s) = 1$ for some pure state. 
From Lemma \ref{lem:inep}, it is an indecomposable and pure effect. 
\hfill $\blacksquare$

In the following, we give a characterization of indecomposable effects in classical, quantum and hyper cuboid systems in order:  

[Classical Systems] Let $\SA_c$ be the state space of a classical system introduced in section \ref{sec:GPT}. 
Remind that any state ${\bm p} = (p_1,\ldots,p_d) \in \SA_c$ has the unique decomposition with respect to pure states: $s = \sum_\mu p_\mu {\bm p}^{(\mu)}$.  
Thus, an effect $e$ on $\SA_c$ is completely characterized by $d$ numbers of value $x_\mu:= e({\bm p}^{(\mu)}) \in [0,1] \ (\mu=1.\ldots,d)$. 
Conversely, for any given $x_\mu \in [0,1] \ (\mu = 1,\ldots,d)$, there exists an effect $e$ such that $e({\bm p}^{(\mu)}) = x_\mu$. 
Let $e^{(\mu)} \in \E(\SA_c) \ (\mu=1,\ldots,d)$ be the effects defined by $e^{(\mu)}({\bm p}^{(\nu)}) = \delta_{\mu \nu}$.
In classical systems, the indecomposable effect is characterized as follows: 
\begin{Prop}\label{prop:CInd}
An effect $e \in \E(\SA_c)$ is indecomposable iff there is one pure state at which the value of effect is non-zero. 
In other words, $\E_{ind}(\SA_c)$ is characterized by $\E_{ind}(\SA_c) = \{\lambda e^{(\mu)} \ | \ \lambda \in (0,1], \mu=1,\ldots,d\}$. 
\end{Prop}
{\bf [Proof] } \ First, let $e$ be an effect such that there exists one pure state, say ${\bm p}^{\mu}$, at which the value of effect is non-zero. Then, one has $e \neq 0$ and $e = \lambda e^{(\mu)} \in \E(\SA_c)$ for some $\lambda \in (0,1]$.  
Let $e = f + g$ for $f,g \in \E(\SA_c)$. 
Then $f({\bm p^{(\nu)}}) = 0 $ for any $\nu \neq \mu$, and it follows that $f = \frac{f({\bm p^{(\mu)}})}{\lambda} e$. 
Therefore, $e$ is indecomposable. 
Next, let $e$ be indecomposable effect. 
Assume that there are at least two non-zero pure states, say ${\bm p^{(\mu_0)}},{\bm p^{(\mu_1)}} \ (\mu_0 \neq \mu_1 = 1,\ldots,d)$ at which the effect values are non-zero. 
Let $x_\mu:= e(\bm p^{(\mu)})$. 
Let $f$, $g$ be effects defined by $f(\bm p^{(\mu)}) = x_{\mu_0} \delta_{\mu \mu_0}$ and $g= e - f$. 
Obviously $e \neq c f$ for any $c \in \R$, and it contradicts that $e$ is indecomposable. 
Since $e \neq 0$, there is the only one pure state at which the value of effect is non-zero. \hfill $\blacksquare$

[Quantum Systems] Next, we show that indecomposable effects for quantum systems are characterized by an one dimensional projections, i.e., rank-one POVM element. 
Let $\HA$ be the $d$ dimensional Hilbert space and let $\SA_q$ be the set of all the density operators on $\HA$. 
We call a non-zero POVM element $E$ indecomposable iff the corresponding effect $e(\cdot) := \tr(E \cdot)$ is indecomposable. 
It is easy to see that a POVM element $E$ is one dimensional iff there exists $\lambda \in (0,1]$ and a unit vector $\psi \in \HA$ such that $E = \lambda \ketbra{\psi}{\psi}$. 
\begin{Prop}\label{prop:charofIndecPOVM}
A POVM element $E \in \E_q$ is indecomposable if and only it is a rank-one POVM element. 
\end{Prop}
{\bf Proof} Let $E = \lambda \ketbra{\psi}{\psi}$ be a rank-one POVM element with a unit vector $\psi \in \HA$ and $\lambda \in (0,1]$. 
Let $E = E_1 + E_2$ for some POVM elements $E_1,E_2$:
\begin{equation}\label{eq:dec1}
\lambda \ketbra{\psi}{\psi} = E_1 + E_2. 
\end{equation} 
Let $\{\psi_n\}_n$ be an orthonormal basis of $\HA$ such that $\psi_1 = \psi$. 
Then, from \eqref{eq:dec1}, it follows that $\bracket{\psi_j}{E_1 \psi_j} = ||E_1^{1/2} \psi_j||^2 =0 \ (\forall j \ge 2)$, and hence $E_1 \psi_j = 0  \ (\forall j \ge 2)$. 
For any $\xi \in \HA$, we have $E_1 \xi = E_1 (\sum_n \bracket{\psi_n}{\xi} \psi_n) = \bracket{\psi}{\xi} E_1 \psi = \ketbra{E_1 \psi}{\psi} \xi$. Thus, $E_1$ has the form of  $\ketbra{\phi}{\psi}$ (where $\phi := E_1 \psi$). Finally, since $E_1$ is Hermitian, it follows that there exists $c^\prime \in \R$ such that $\phi = c^\prime \psi$ and hence $E_1 = c^\prime \ketbra{\psi}{\psi} = c E$ where $c := \frac{c^\prime}{\lambda}$. This implies that $E$ is indecomposable. 
Next, let $E$ be indecomposable. 
Assume that $E$ is rank $l$ POVM element for some $l \ge 2$, and let 
$E = \sum_{n=1}^l c_n \ketbra{\psi_n}{\psi_n} \ (c_n \in (0,1])$ be an eigenvalue decomposition of $E$. 
Let $E_1 := c_1 \ketbra{\psi_1}{\psi_1}$ and $E_2 := \sum_{n=2}^l c_n \ketbra{\psi_n}{\psi_n} $. Obviously, they are POVM elements satisfying $E = E_1 + E_2$. 
However, for any $c \in \R$, we have $E \neq c E_1$ (For instance, $E \psi_2 = c_2 \neq 0$ while $c E_1 \psi_2 = 0$). 
This contradicts that $E$ is indecomposable. 
Since $E \neq 0$, we conclude that $E$ is rank one POVM element. 
 $\blacksquare$

[Hyper cuboid systems] Finally, let $\SA_{\mathrm{cb}}$ be the state space of a $d$ dimensional hyper cuboid system introduced in section \ref{sec:GPT}. 
To determine the indecomposable effects in $\SA_{\mathrm{cb}}$, we present a general lemma which is also useful in later arguments (see Appendix \ref{app:proofs} for the proof):
\begin{Lem}\label{lem:Eindtakes0}
If the state space $\SA$ of a GPT contains at least two states, then for every indecomposable effect $e \in \E(\SA)$ we have $e(s) = 0$ for some $s \in \SA$.
\end{Lem}
By virtue of this lemma, we obtain the following characterization of indecomposable effects in $\SA_{\mathrm{cb}}$:
\begin{Prop}\label{prop:charofIndecforHCB}
An effect $e \in \E(\SA_{\mathrm{cb}})$ is indecomposable if and only if it is nonzero and it takes $0$ at a $d-1$ dimensional face (facet) of $\SA_{\mathrm{cb}}$.
\end{Prop}
{\bf Proof} First we consider the \lq if' part.
Suppose that $e$ is nonzero and $e$ takes $0$ at a facet $F$ of $\SA_{\mathrm{cb}}$.
Fix a state $s \in \SA_{\mathrm{cb}}$ such that $s \not\in F$.
Note that $e(s) > 0$ since $e$ is nonzero.
If $e$ decomposes as $e = e_1 + e_2$ with $e_1,e_2 \in \E(\SA_{\mathrm{cb}})$, then $e_1$ (hence $e_2$) also takes $0$ at $F$.
This implies that $e_1 = \lambda e$ where $\lambda = e_1(s) / e(s) \in \R$, hence $e$ is indecomposable.

Second, we consider the \lq\lq only if'' part.
By Lemma \ref{lem:Eindtakes0}, an indecomposable effect $e$ takes $0$ at some state, hence at some pure state in $\SA_{\mathrm{cb}}$.
By symmetry, we may assume without loss of generality that $e(s_0) = 0$ where $s_0 = (0,0,\dots,0) \in \SA_{\mathrm{cb}}$.
Let $s_i$ ($i \in \{1,2,\dots,d\}$) be the vertex of $\SA_{\mathrm{cb}}$ such that its $j$-th component is $\delta_{i j}$.
Then we have $e = \sum_{i = 1}^{d} e(s_i) e_i$ where $e_i \in \E(\SA_{\mathrm{cb}})$ maps $(c_1,c_2,\dots,c_d)$ to $c_i$.
Since $e$ is indecomposable, it follows that $e = \lambda e_i$ for some $1 \leq i \leq d$ and $\lambda \in \R$, therefore $e$ takes $0$ at the facet $\{(c_1,\dots,c_d) \in \SA_{\mathrm{cb}} \mid c_i = 0\}$ of $\SA_{\mathrm{cb}}$.
\hfill $\blacksquare$

For example, the indecomposable effects in the squared system (i.e., when $d = 2$) are listed in Table \ref{tab:effects_square_space}, where $\alpha_1,\dots,\alpha_4 \in \left(0,1\right]$ are parameters.
\begin{table}[htb]
\centering
\caption{Indecomposable effects in $\SA_{\mathrm{cb}}$, $d = 2$}
\label{tab:effects_square_space}
\begin{tabular}{|c||c|c|c|c|} \hline
& \multicolumn{4}{|c|}{value at} \\
effect & $(0,0)$ & $(0,1)$ & $(1,0)$ & $(1,1)$ \\ \hline\hline
$e_1$ & $0$ & $0$ & $\alpha_1$ & $\alpha_1$ \\ \hline
$e_2$ & $\alpha_2$ & $\alpha_2$ & $0$ & $0$ \\ \hline
$e_3$ & $0$ & $\alpha_3$ & $0$ & $\alpha_3$ \\ \hline
$e_4$ & $\alpha_4$ & $0$ & $\alpha_4$ & $0$ \\ \hline
\end{tabular}
\end{table}

\subsection{Indecomposable and complete measurements}\label{subsec:IndCom}

Using the indecomposable effects defined above, we define an indecomposable measurement in any GPT as follows: 
\begin{Def}
In a GPT with a state space $\SA$, we say that a measurement $M = (m_j)_j$ is indecomposable if all $m_j \in \E(\SA)$ are indecomposable. 
The set of all the indecomposable measurements is denoted by $\M_{\mathrm{ind}}(\SA)$ or simply by $\M_{\mathrm{ind}}$.
\end{Def}
From Proposition \ref{prop:charofIndecPOVM}, an indecomposable measurement is a generalization of a one-rank POVM measurement in quantum systems. 
\begin{Prop}
In any GPT, there exists an indecomposable measurement, i.e., 
$\M_{\mathrm{ind}}(\SA) \neq \emptyset$. 
\end{Prop}
{\bf Proof} \ 
From Lemma \ref{lem:EdecInd}, a decomposition of the unit effect $u$ with respect to the indecomposable effects gives an indecomposable measurement. 
\hfill $\blacksquare$

In quantum systems, rank-one PVM measurement plays a fundamental role in the foundation of quantum physics, which describes a measurement of a non degenerate Hermitian operator. 
One can also define the correspondent notion in any GPT, which we call a {\it complete} measurement, as follows: 
\begin{Def}
In a GPT with a state space $\SA$, we say that a measurement $M = (m_j)_j$ is complete if all $m_j \in \E(\SA)$ are indecomposable and extreme. 
The set of all the complete measurements are denoted by $\M_{\mathrm{comp}}(\SA)$, or simply by $\M_{\mathrm{comp}}$. 
\end{Def}
It is easy to see that the set of extreme effects for classical systems are characterized by $\E(\SA_c)_{\mathrm{ex}}= \{e \in \E(\SA_c) \ | \ 
e(\bm p^{(i)}) = 1\  \mathrm{or} \ 0 \ (i =1,\ldots,d)\}$ \footnote{Let $e $ be an effect where $e(\bm p^{(i)}) = 1\  \mathrm{or} \ 0$ for any $i$. 
Assume that there exists $f,g \in \E(\SA_c)$ and $\lambda \in (0,1)$ such that 
$e = \lambda f + (1-\lambda) g$. 
Since $e(\bm p^{(i)})$ are $1$ or $0$ and $\lambda \in (0,1)$, 
one can show $f(\bm p^{(i)})$ are also restricted to be $1$ or $0$. 
Therefore, $e = f = g$, and $e$ is an extreme point. 
Next, let $e$ be an effect where there exists $i_0$ such that $e(\bm p^{(i_0)}) \in (0,1)$. 
Let $i_m \in \{1,\ldots,d\}$ such that $e(\bm p^{(i_m)})$ is a minimum value among all $e(\bm p^{(i)}) \neq 0$, i.e., 
$e(\bm p^{(i)}) \neq 0 \Rightarrow e(\bm p^{(i)}) \ge e(\bm p^{(i_m)})$. 
Let $f,g$ be effects defined by $f (\bm p^{(i)})= \delta_{i i_m}$, i.e., $f = e^{(i_m)}$ and $g(\bm p^{(i)})= \frac{x_i-x_{i_m}}{1-x_{i_m}}$. It is easy to see $e \neq f,g$ and $e = \lambda f + (1-\lambda) g$ for $\lambda := e(\bm p^{(i_m)}) \in (0,1)$. Therefore, $e $ is not an extreme point and this completes the proof. $\blacksquare$}. 
Hence, from proposition \ref{prop:CInd}, there is essentially the unique complete measurement in classical systems given by $M_\mathrm{comp} := (e^{(j)})_j$, where $e^{j}({\bm p}^{(k)}):= \delta_{jk} \ (j,k=1,\ldots,d)$. 
More precisely, $M = (m_j)_j$ is a complete measurement iff $m_j = e^{\sigma(j)}$ where $\sigma(j)$ is a permutation of $(1,\ldots,d)$. 
On the other hand, in quantum systems, a complete measurement is given by a rank-one PVM measurement. This follows from Proposition \ref{prop:charofIndecPOVM} and the fact that a POVM element is extreme iff it is a projection operator. 

By definition, $\M_{\mathrm{comp}}(\SA) \subset \M_{\mathrm{ind}}(\SA)$. 
However, the existence of the complete measurements does not necessarily hold for any GPT (See Appendix \ref{app:AppB} for a counter example).

\section{Some quantities related to an entropy}

In this section, we consider three quantities on $\SA$ in any GPT which are related to the notion of entropy. 
Indeed, all of them coincides with the Shannon entropy $H$ and von Neumann entropy $S$ in classical and quantum systems, respectively, and therefore give generalizations of entropies in classical and quantum systems. 
However, as is shown, they do not coincide in some GPTs, and does not satisfy some of properties of an entropy. 
In the following, let $H({\bm p})$, or simply as $H(p_i)$, denote the Shannon entropy for a probability distribution ${\bm p} = (p_1,\ldots,p_d)$: 
$H({\bm p}) := - \sum_i p_i \log p_i$. 
We also denote it by $H(X)$ when the random variable $X$ are dealt with. 
The mutual information for a random variable $X$ and $J$ are denoted by 
$H(X:J) := H(X) + H(J) - H(X,J)$.  
In quantum systems, the von Neumann entropy for a density operator $\rho$ on $\HA$ is denoted by $S(\rho) : = - \tr (\rho \log \rho) $. 

Let us consider a general GPT with a state space $\SA$. 
For any state $s \in \SA$, we denote by $\D(s)$ the set of all the ensembles
$\{p_x;s_x\}_x \ (s_x \in \SA, p_x \ge 0, \sum_x p_x = 1)$ such that $s = \sum_x p_x s_x$. 
The set of all the ensembles for $s$ with respect to pure states are denoted by $\PA(s) \subset \D(s)$; i.e., $\{p_x;s_x\} \in \PA(\SA) \Leftrightarrow s = \sum_x p_x s_x, \ s_x \in \SA_{pure}$. 
Note that $H$ and $S$ have a concavity property. 
Both $H$ and $S$ are positive and take the minimum value $0$ iff the state is pure.   
The following upper bound of von Neumann entropy is also well known: for a probability distribution $(p_i)_i$ and a set of density operators $\{\rho_i\}_i$, 
\begin{equation}\label{eq:UPofVNE}
S(\sum_i p_i \rho_i)  \le H(p_i) + \sum_i p_i S(\rho_i), 
\end{equation}
with equality iff density operators $\rho_i$ are orthogonal to each other. 
See, for instance \cite{ref:NC}, for the properties of Shannon and von Neumann entropies.

In any GPT, let us define the following quantities for $s \in \SA$: 
\begin{eqnarray} 
S_1(s) &:=& \inf_{M = (m_j)_j \in \M_{\mathrm{ind}}} H(m_j(s)). \label{eq:S1}\\
S_2(s) &:=& \sup_{\{p_x,s_x\} \in {\cal P}(s) }\sup_{M = (m_j)_j \in \M_{\mathrm{ind}}} H(X:J). \label{eq:S2} \\
S_3(s) &:=& \inf_{\{p_x,s_x\} \in {\cal P}(s)} H(p_x). \label{eq:S3} 
\end{eqnarray}
In $S_2(s)$, $H(X:J)$ is defined by a joint distribution $p_x m_j(s_x)$ with an ensemble $\{p_x,s_x\} \in \PA(s)$ and a measurement $M=(m_j)_j$. 
From the definition and the positivity of the Shannon entropy and the mutual information, the positivity of $S_1,S_2,S_3$ are obvious. 
It is easy to see that $S_2$ can be redefined with respect to $\D(s)$ and $\M$: 
\begin{Lem}\label{lem:S3} We have 
$$
S_2(s) = \sup_{\{p_x,s_x\} \in \D(s) }\sup_{M = (m_j)_j \in \M} H(X:J)
$$
\end{Lem}
{\bf Proof} \ A straightforward computation shows that, for any $\{p_x,s_x\} \in \D(s)$ and $M = (m_j)_j \in \M$,
the value of $H(X:J)$ is not decreased by replacing $\{p_x,s_x\}$ with the pure state decomposition
of $s$ obtained by decomposing every $s_x$ into pure states, and by replacing $M$ with the
indecomposable measurement obtained by decomposing every $m_j$ into indecomposable effects (cf. Proposition \ref{lem:EdecInd}).
This implies the desired relation.
\hfill $\blacksquare$

However, note that it is essential to use $\M_{\mathrm{ind}}$ and $\PA(s)$ for the definitions of $S_1$ and $S_3$. 
Indeed, redefinitions of $S_1$ and $S_3$ with respect to $\D(s)$ and $\M$ give trivial quantities: $\inf_{M \in \M} H(m_j(s)) = 0, \ \inf_{\{p_x,s_x\} \in {\cal D}(s)} H(\{p_x\}_x) =0$. 

Notice that all three quantities \eqref{eq:S1}-\eqref{eq:S3} are defined with physical languages:  
$S_1(s)$ measures the minimum uncertainty of measurement among indecomposable measurements under a state $s$; 
$S_2(s)$ measures the maximum accessible information (by an optimal measurement) among any preparation of $s$ (See below). 
Finally, $S_3(s)$ measures the minimum uncertainty for a preparation of $s$ with respect to pure states. 

Indeed, under the preparation of states $s_x$ with a prior probability distribution $p_x$, the accessible information $I(\{p_x,s_x\})$ is defined by $\sup_{M = (m_j)_j \in \M} H(X:J)$ where the joint probability distribution between $X$ and $J$ (measurement outcome by a measurement $M = (m_j)_j$ ) is given by $p(x,j):= p_x m_j(s_x)$. 
Therefore, from Lemma \ref{lem:S3}, we have $S_2(s) = \sup_{\{p_x,s_x\} \in \D(s)} I(\{p_x,s_x\})$, and thus 
\begin{Prop}
In GPT, for any preparation of states $\{p_x,s_x\}$, the accessible information is bounded as
\begin{equation}\label{eq:UPofI}
I(\{p_x,s_x\}) \le S_2(s),
\end{equation}
where $s := \sum_x p_x s_x$. 
\end{Prop}

Notice that, in quantum systems, the Holevo bound \cite{ref:HB} gives an upper bound of the accessible information by the Holevo $\chi$ quantity: 
For a preparation of density operators $\rho_x$ with a probability distribution $p_x$, 
\begin{equation}\label{eq:HB}
I(\{p_x,\rho_x\}) \le \chi:= S(\rho) - \sum_x p_x S(\rho_x). 
\end{equation} 
In the following, we see that $S_2$ coincides with the von Neumann entropy in quantum systems. 
Thus, \eqref{eq:UPofI} gives a looser bound than the Holevo bound in quantum systems. 
(For the pure state ensemble, \eqref{eq:UPofI} gives exactly the Holevo bound since the von Neumann entropy vanishes on pure states.)  

Now, we show that all three quantities \eqref{eq:S1}-\eqref{eq:S3} are generalizations of Shannon and von Neumann entropies in classical and quantum systems: 
\begin{Thm} $\mathrm{(i)}$ In classical systems, $S_1(s),S_2(s),S_3(s) $ are the Shannon entropy. 
$\mathrm{(ii)}$ In quantum systems, $S_1(s),S_2(s),S_3(s)$ are the von Neumann entropy.  

\end{Thm}
{\bf Proof} \ 
(i) Let $\SA_c$ be the state space of a classical system. 
From Proposition \ref{prop:CInd}, any indecomposable measurement in classical system is given by $(\lambda_{i,\mu} e^{(\mu)})_{i,\mu}$ where 
$\lambda_{i,\mu} \ge 0, \sum_i \lambda_{i,\mu} = 1$ for any $\mu = 1,\ldots,d$. 
Thus, for a state ${\bm p} = (p_1,\ldots,p_d) \in \SA_c$, the probability distribution given by the indecomposable measurement is $(\lambda_{i,\mu} p_\mu)_{i,\mu}$. 
Note that from the concavity of the function $g(x):= -x \log x \ (x\in [0,1])$ with the convention $g(0)=0$, it holds that $g(\lambda x) \ge \lambda g(x) $, and thus we have $H(\lambda_{i,\mu} p_\mu) =  \sum_{i,\mu} g(\lambda_{i,\mu} p_\mu) \ge \sum_\mu (\sum_i \lambda_{i,\mu}) g(p_\mu) = H(p_\mu)$. 
Thus, we have $S_1({\bm p}) \ge H({\bm p})$. 
Since $(e^{(\mu)})_\mu$ is an indecomposable measurement with which the probability distribution is given by ${\bm p}$, we have $S_1({\bm p}) = H({\bm p})$. 

As mentioned before, a state space of a classical system is characterized by a simplex. 
Thus, we have 
$$
S_2({\bm p}) = \sup_{M \in \M} H(X:J) 
$$
where the random variable $X$ is described by the probability distribution 
${\bm p}$. 
Remind that the mutual information can be written as $H(X:J) = H(X) - H(X|J)$
 where $H(X|J)$ denotes the conditional entropy, and it follows that 
$$
S_2({\bm p}) = \sup_{M \in \M} H(X:J) = H(X) - \inf_{M \in \M} H(X|J). 
$$
Since there exists a measurement $M=(m_j)$ to discriminate all pure states in a classical system, we have $\inf_{M \in \M} H(X|J) = 0$ (i.e., the uncertainty of $X$ conditioned on the information of $J$ is zero). 
Therefore, we have $S_2({\bm p}) = H(X) = H({\bm p})$. 

Again from the unique pure state decomposition,  
there exists the unique ensemble $\{p_\mu,p^{(\mu)}\}_{\mu=1}^d$ for any state ${\bm p} = (p_1,\ldots,p_d) \in \SA_c$. 
Therefore, we have $S_3({\bm p}) = H({\bm p})$. 

(ii) Next, we consider a quantum system described by a Hilbert space $\HA$. 
First, let $f$ be a concave function on $[0,1]$ such that $f(0)=0$, and let $\rho$ be a density operator on $\HA$. Then, it is easy to show 
\footnote{Let $\rho = \sum_{j=1}^d \rho_j \ketbra{\phi_j}{\phi_j}$ be an eigenvalue decomposition of $\rho$. 
Notice that $\{\phi_j\}_{j=1}^d$ is an orthonormal basis of $\HA$ and $\sum_{j=1}^d |\bracket{\psi}{\rho_j}|^2 = ||\psi||^2 \le 1$. Thus $(q_j)_{j=1}^{d+1}$ where $q_j := |\bracket{\psi}{\rho_j}|^2 \ (j=1,\ldots,d)$ and $q_{d+1} := 1- \sum_j |\bracket{\psi}{\rho_j}|^2$ is a probability distribution. 
From the concavity of $f$, we get 
$$
f(\bra{\psi}\rho\ket{\psi}) = f(\sum_{j=1}^{d+1} \rho_j q_j ) \ge \sum_{j=1}^{d+1} q_j f(\rho_j) = \bra{\psi}f(\rho)\ket{\psi}, 
$$
where $\rho_{d+1}:= 0$. 
} that for all vector $\psi \in \HA $ such that $||\psi|| \le 1$, we have 
$$
f(\bra{\psi}\rho\ket{\psi}) \ge \bra{\psi}f(\rho)\ket{\psi}.
$$

Let us fix any indecomposable POVM measurement $(E_j)_j$ on quantum system $\HA$, i.e., rank-one POVM measurement. 
We can write $E_j = \ketbra{\psi_j}{\psi_j}$ with a vector $\psi \in \HA$ such that $0 <||\psi_j|| \le 1$ and $\sum_i \ketbra{\psi_j}{\psi_j} = \I$. 
Remind that the von Neumann entropy of $\rho$ is defined by 
$$
S(\rho) := \tr g(\rho) 
$$
with the concave function $g(x):= -x \log x$ with the convention $g(0) = 0$. 
Applying $g$ to the above concave function $f$, we have 
$
H(e_j(\rho)) = \sum_{j} g(\bra{\psi_j}\rho\ket{\psi_j}) \ge \sum_{j} \bra{\psi_j}g(\rho)\ket{\psi_j} 
= \tr (g(\rho) \sum_i E_i)= \tr g(\rho) = S(\rho)$. 
By considering the indecomposable measurement given by $(\ketbra{\phi_j}{\phi_j})_j$ where $\phi_j$s are complete eigenvectors of $\rho$, we obtain $S_1(\rho) = S(\rho)$. 

Next, from the Holevo bound \eqref{eq:HB}, we have 
$$
S_2(\rho) \le S(\rho) - \inf_{\{p_x,\rho_x\} \in {\cal D}(\rho)} (\sum_x p_x S(\rho_x)) = S(\rho). 
$$
The final equality follows from the eigenvalue decomposition $\rho = \sum_x p_x \ketbra{\phi_x}{\phi_x}$ and $S(\ketbra{\phi_x}{\phi_x}) = 0$. 
Again with the decomposition $\{p_x,\rho_x = \ketbra{\phi_x}{\phi_x}\}$ of eigenvalues and eigenvectors, 
there exists an optimal measurement $M_j:= \ketbra{\phi_j}{\phi_j}$ to discriminate $\rho_x$, and thus one has $H(X:J) = H({\bm p})$. 
Since $S(\rho) = H({\bm p})$, we have $S(\rho) = H(X:J) \le S_2(\rho)$. 

Finally, let $\{p_x,\rho_x\} \in \PA(\rho)$ be a pure state decomposition of $\rho$. 
Then, from the inequality \eqref{eq:UPofVNE} and the fact that $S(\rho_x) = 0$ for pure states $\rho_x$, we have 
$$
S(\rho) \le H(\{p_x\}) \le S_3(\rho). 
$$ 
Moreover, an eigenvalue decomposition $\rho = \sum_x p_x \ketbra{\phi_x}{\phi_x}$ of $\rho$ gives a pure state decomposition such that $\rho_x = \ketbra{\phi_x}{\phi_x}$ are orthogonal to each other, we have the equality: $S_3(\rho) = S(\rho)$. This completes the proof. \hfill $\blacksquare$ 

Notice that the fact that $S_1,S_2,S_3$ coincide with the von Neumann entropy in quantum systems shows that we have alternative expressions with operational meanings for the von Neumann entropy. 
The characterization of $S$ by $S_3$ has been noticed by Jaynes \cite{ref:Jaynes}. 
Here, we remark that $S_1$ could be defined by the infimum of Shannon entropy among not indecomposable measurements but complete measurements. 
Then, it is easy to restate the above mentioned proof to show $S_1$ coincides with Shannon and von Neumann entropy in classical and quantum systems. 
However, as we have noticed in Sec.~\ref{subsec:IndCom}, there exists a GPT where no complete measurements exists. 
This is the reason why we have defined $S_1$ among indecomposable measurements. 
 
In order to see the properties of $S_1,S_2,S_3$ in a general GPT, let us again consider the squared system $\SA_{\mathrm{sq}}$. 
Let $h(x) := -x \log x - (1-x) \log (1-x)$ is the binary Shannon entropy:  
\begin{Prop}\label{prop:entropy_sq}
In the squared system, for $s = (c_1,c_2) \in \SA_{\mathrm{sq}}$, we have 
\begin{equation}\label{eq:S1sq}
S_1(s) = \min[h(c_1),h(c_2)], 
\end{equation}
\begin{equation}\label{eq:S2sq}
S_2(s) = \max[h(c_1),h(c_2)], 
\end{equation} 
\begin{eqnarray}\label{eq:S3sq}
S_3(s) = \left\{
\begin{array}{cc}
k(c_1+c_2-1) & s \in R_{1L} \\
k(c_1) & s \in R_{1R} \\
k(0) & s \in R_{2U} \\
k(c_1) & s \in R_{2B} \\
k(c_2) & s \in R_{3U} \\
k(c_1+c_2-1) & s \in R_{3B} \\
k(c_2) & s \in R_{4L} \\
k(0) & s \in R_{4R} 
\end{array}
\right. 
\end{eqnarray}
where $k(x) := H(\{x,c_1-x,c_2-x,1+x-c_1-c_2\}) $ and the regions $R_{1L},\cdots,R_{4R} \subset \SA_{\mathrm{sq}}$ are given in Fig.~\ref{fig:sq}-$\mathrm{(4)}$. 
\end{Prop}
(See Appendix A for the proof.) See graphs of $S_1,S_2$ and $S_3$ in Fig.~\ref{fig:sq}-(1)-(3). 
Moreover, in $\SA_{\mathrm{sq}}$, the following relations among $S_1,S_2$ and $S_3$ are true: 
\begin{Prop}\label{prop:relationofS} For any $s \in \SA_{\mathrm{sq}}$, 
$$
S_1(s) \le S_2(s) \le S_3(s)  
$$
\end{Prop}
\begin{figure} 
\includegraphics[height=0.3 \textwidth]{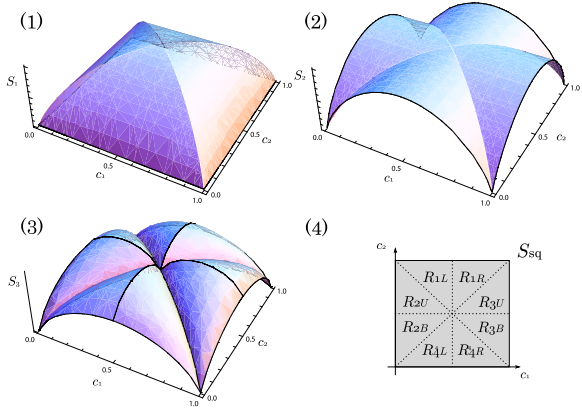}
\caption{In the squared systems, (1), (2) and (3) show the graphs of $S_1$, $S_2$ and $S_3$. (4) specifies the region $R_{1L},\ldots,R_{4R}$.}\label{fig:sq}
\end{figure}
(See Appendix A for the proof.)


\subsection{Concavity}

In this section, we consider the concavity properties of $S_1,S_2$ and $S_3$. 
It turns out that $S_1$ is concave on $\SA$ in any GPT, while there exists  GPT models where $S_2$ and $S_3$ are not concave. 
\begin{Prop}
In any GPT, $S_1$ is concave on $\SA$. 
\end{Prop} 
{\bf Proof} \ 
Let $(p_x)_x \ (x=1,\ldots,m)$ be a probability distribution and let $s_x \in \SA \ (x=1,\ldots,m)$. 
Then, from the affinity of effects $m_j$ and the concavity of the Shannon entropy, we have  
\begin{eqnarray}
S_1(\sum_x p_x s_x) &=& \inf_{M = (m_j)_j \in \M_{\mathrm{ind}}} H(e_j(\sum_x p_x s_x)) \nonumber \\
&\ge & \inf_{M = (m_j)_j \in \M_{\mathrm{ind}}} \sum_x p_x  H( m_j(s_x)) \nonumber \\
&\ge & \sum_x p_x  \inf_{M = (m_j)_j \in {\cal M}_{\mathrm{ind}}}  H(m_j(s_x)) \nonumber \\
&=& \sum_x p_x S_1(s_x).
\end{eqnarray}
\hfill $\blacksquare$

Contrast to $S_1$, $S_2$ and $S_3$ does not satisfy the concavity in some GPT. 
It is easy to give counter examples but it is obvious that concavity does not hold in the squared systems from the Fig.~\ref{fig:sq}-(2) and (3). 

In stead of the concavity, we show the following: $S_2$ satisfies the following {\it weak concavities}: 
\begin{Prop}(Weak Concavity) 
In any GPT, $S_2$ satisfies the followings: 
\begin{eqnarray}
S_2(\sum_x p_x s_x) \ge \frac{\sum_x p^2_x S^2_2(s_x)}{\sum_x p_x S_2(s_x)}, \label{eq:WeakConcavity1} \\
S_2(\sum_x p_x s_x) \ge \sum_x p^2_x S_2(s_x), \label{eq:WeakConcavity2} \\
S_2(\sum_x p_x s_x) \ge \frac{1}{|X|}\sum_x p_x S_2(s_x), \label{eq:WeakConcavity3} \\
S_2(\sum_x p_x s_x) \ge \max_x p_x S_2(s_x), \label{eq:WeakConcavity4}
\end{eqnarray}
for any $\{p_x,s_x\}_{x \in X} \in \D(s)$ (in (\ref{eq:WeakConcavity1}), we interpret the right-hand side as $0$ if $S_2(p_x)$ are all $0$). 
\end{Prop}
{\bf Proof} \ 
To prove this proposition, we use the following lemma (see Appendix \ref{app:proofs} for the proof):
\begin{Lem}
\label{lem:lemma_weak_concavity}
Let $\{p_x,s_x\}_{x \in X} \in \D(s)$, $p_x \neq 0$.
Then for any value $\pi_x \geq 0$, $x \in X$ such that $\sum_{x} \pi_x = 1$, we have
\begin{displaymath}
S_2(\sum_{x} p_x s_x) \geq \sum_{x} \pi_x p_x S_2(s_x) \enspace.
\end{displaymath}
\end{Lem}
By using this lemma, (\ref{eq:WeakConcavity1}) is proved by putting $\pi_x = p_x S_2(s_x) / (\sum_{x'} p_{x'} S_2(s_{x'}))$ (here we may assume that the denominator of the $\pi_x$ is nonzero, as otherwise the claim is obvious); (\ref{eq:WeakConcavity2}) is proved by putting $\pi_x = p_x$; (\ref{eq:WeakConcavity3}) is proved by putting $\pi_x = 1 / |X|$; and (\ref{eq:WeakConcavity4}) is proved by putting $\pi_x = \delta_{x x_0}$ where $x_0 \in X$ is such that $p_{x_0} S_2(s_{x_0}) = \max_x p_x S_2(s_x)$.
\hfill $\blacksquare$


\begin{Prop} 
In any GPT, $S_3$ satisfies 
$$
S_3(\sum_{i=1}^m p_i s_i) \le H(p_i) + \sum_{i=1}^m p_i S_3(s_i), 
$$
for any $s_i \in \SA \ (i=1,\ldots,m)$ and probability distribution $(p_i)_{i=1}^m$ . 
\end{Prop}
{\bf Proof} 
Let $\epsilon > 0$ be an arbitrary positive number. 
Let $s := \sum_{i=1}^m p_i s_i$ and let $\{p^i_j;s^i_j\}_j \in \PA(s_i)$ be an ``optimal" decomposition of $s_i$ such that $S_3(s_i) + \epsilon \ge H(p^i_j)$. 
Since $\{p_i p^i_j\}_{i,j} \in \PA(s)$, we have 
$S_3(s) \le H(I,J) := - \sum_{i,j} p_i p^i_j \log(p_i p^i_j) = -\sum_i p_i \log p_i + \sum_i p_i (-\sum_j p^i_j \log p^i_j) = H(p_x) + \sum_i p_i H(p^i_j) \le H(p_x) + \sum_i p_i S_3(s_i) + \epsilon$. \hfill $\blacksquare$ 

Thus, $S_3$ satisfies the same upper bound \eqref{eq:UPofVNE} of the von Neumann entropy in any GPT.

\subsection{Measure for pureness}

Since both the Shannon entropy and the von Neumann entropy vanishes if and only if the state is pure, they can be considered as a measure of pureness. 
(Note also that they take the maximum value iff the state is the maximal mixed states.) 
We show that $S_2$ and $S_3$ has this desired property in any GPT, while  $S_1$ does not satisfy this in general. 
\begin{Prop} In any GPT, $\mathrm{(i)}$ $S_2(s)  = 0$ if and only if $s$ is pure. 
$\mathrm{(ii)}$ $S_3(s)  = 0$ if and only if $s$ is pure. 
\end{Prop}
{\bf Proof} 
 (i) Let $s \in \SA$ be a pure state. Since $s$ is an extreme point of $\SA$, 
$\PA(\rho)$ has essentially the unique (trivial) decomposition: $\{1;s\}$ with $H(X) = 0$. 
Thus, we have 
$$
S_2(s) = \sup_{M=(m_j)_j \in \M} H(X:J) = - \inf_M H(X|J) \le 0.   
$$ 
To see the converse, let $S(s) = 0$ for $s \in \SA$ and let $s = p_1 s_1 + p_2 s_2$ where $p_1 \in (0,1), p_1 + p_2 = 1$ and $s_1,s_2 \in {\cal S}$. 
Since $S_2(s) = 0$ and $\{p_x;s_x\}_{x=1,2} \in D(s)$, we have 
$$
H(X:J) = 0
$$ 
for the random variable $X=1,2$ and for any $M=(m_j)\in {\cal M}$. 
This implies that the joint probability $p(x,j):= p_x m_j(s_x)$ is a product state, or equivalently, the conditional probability $p(j|x):= p(x,j)/p_x = m_j(s_x)$ is independent of $x$ (Notice that $p_1,p_2 \neq 0$). 
In particular, we have $m_j(s_1) = p(j|1) = p(j|2) = m_j(s_2)$. 
Since this folds for any effect $m_j$, we have $s_1 = s_2$ from the separating property of states. 
Therefore, $s$ has only the trivial decomposition and is a pure state.  

(ii) 
Let $s \in \SA$ be a pure state and thus $\PA(s)$ has essentially the unique (trivial) decomposition: $\{1;s\}$ where $H(X) = 0$. 
Thus, we have $S_3(s) = H(X) = 0$. 
Conversely, let $S_3(s) = 0$. Then, for any $\{p_x;s_x\}_x \in \PA(\SA)$, it follows $H(X) = 0$. 
Assume that $s$ is not a pure state. 
Then, we have $\{p_x,s_x\}_{X} \in \PA(\SA)$ where $p_{x_1},p_{x_2} > 0 $ for some $x_1,x_2 \in X$. 
However, this contradicts that $H(X) = 0$. 
Therefore, $s$ is a pure state.

\hfill $\blacksquare$

Contrast to $S_2$ and $S_3$, $S_1$ does not satisfy this property. 
For instance, from \eqref{eq:S1sq}, $S_1(s) = 0$ for any state $s$ on the boundary (four edges) of $\SA_{\mathrm{sq}}$. 
(See Fig.~\ref{fig:sq}-(1). 
Note that $s$ on edges but not on vertexes is not a pure state.) 
In general GPT, we show the followings: 
\begin{Prop}\label{prop:S1andpseudoclassical}
In any GPT, $S_1(s) = 0$ implies that $s$ is on the boundary of $\SA$.
\end{Prop}
{\bf Proof} \ 
It suffices to consider the case that $\SA$ has at least two states.
To prove this proposition, we use the following two lemmas (see Appendix \ref{app:proofs} for the proofs):
\begin{Lem}
\label{lem:S1andpseudoclassical_1}
Let $k,\ell \geq 1$ be an integer.
Let $h(x) = -x \log x$.
If $x_1,\dots,x_{\ell} \in [0,1/k]$ and $\sum_j x_j = 1$, then $H(x) = \sum_j h(x_j) \geq \log k$.
\end{Lem}
\begin{Lem}
\label{lem:S1andpseudoclassical_2}
For any $s \in \SA$, the map $f_s:\E(\SA) \times \SA \to \R$, $f_s(e,t) = e(s) - e(t)$, is continuous.
\end{Lem}
Let $s \in \SA$ such that $S_1(s) = 0$.
First we show that $\sup_{(e,t)} f_s(e,t) = 1$.
Let $k \geq 2$ be any integer.
Since $S_1(s) = 0$, there is an indecomposable measurement $M = (m_i)_i \in \M_{\mathrm{ind}}$ such that $H(m_i(s)) < h(1/k)$ ($< \log k$).
Then we have $m_i(s) \geq 1 - 1/k$ for some $i$, as otherwise we have a contradiction as follows:
If $m_{i_0}(s) \in (1/k,1-1/k)$ for some $i_0$ then we have $H(m_i(s)) \geq h(m_{i_0}(s)) > h(1/k)$; while if $m_i(s) \leq 1/k$ for all $i$ then we have $H(m_i(s)) \geq \log k$ by Lemma \ref{lem:S1andpseudoclassical_1}.
For this $m_i$, Lemma \ref{lem:Eindtakes0} implies that there is a state $t \in \SA$ such that $m_i(t) = 0$.
This implies that $f_s(m_i,t) \geq 1 - 1/k$.
Since $k \geq 2$ is arbitrary, we have $\sup_{(e,t)} f_s(e,t) = 1$.
Since $\E(\SA) \times \SA$ is compact, Lemma \ref{lem:S1andpseudoclassical_2} implies that $f_s(e,t) = 1$ for some $e \in \E(\SA)$ and $t \in \SA$, therefore $e(s) = 1$ and $e(t) = 0$.
This implies that $e$ is not constant and $s$ lies in a supporting hyperplane of $\SA$, hence $s$ is on the boundary of $\SA$ as desired.
\hfill $\blacksquare$

Note that, in $\SA_{\mathrm{sq}}$, the converse is also true: all states on the boundary $s$ satisfy $S_1(s) = 0$. 
However, this is not the case for any GPT. 
In particular, one can construct a GPT where $S_1(s) \neq 0$ even for a pure state $s$. 
For instance, consider a GPT introduced in Appendix \ref{app:AppB} with state space $\SA \subset \R^2$, which has the four pure states $(0,0)$, $(1,0)$, $(0,1)$, and $(2,2)$.
Then any indecomposable effect in $\SA$ is of the form $\lambda e_i$ such that $0 < \lambda \leq 1$ and $e_i$ is one of the four effects in Table \ref{tab:effects_skew_square_space} in Appendix \ref{app:AppB}.
This implies that for any indecomposable measurement $M = (m_i)_i \in \M_{\mathrm{ind}}$ we have $m_i(0,0) \leq 2/3$ for all $i$, therefore $S_1(0,0) > 0$ (see Lemma \ref{lem:S1andpseudoclassical_1}).
Thus, in general GPT, neither directions of ``$s$ is pure $\Leftrightarrow$ $S_1(s) = 0$'' does not holds in general. 
In the next section, we consider a class of GPTs with fairly fine property.  

\section{Principle of Equality for Pure States and Symmetric GPT} 

In the last part of the preceding section, we considered a GPT where for some pure state $s$ it holds that $S_1(s) > 0$ (See GPT in Appendix B). 
However, the structure of state space are asymmetric and might be just toy models for GPTs. 
On the other hand, both classical and quantum systems has a certain class of symmetric structures: 
In particular, there are no special pure states which have different properties from another pure states. 
We call this the principle of {\it equality for pure states} and can be formulated as follows:
\begin{Def}(Equality for pure states)
We say that GPT satisfy the principle of equality for pure states if,  for any pure states $s_1,s_2 \in \SA$, there exists a bijective affine map $f$ on $\SA$ such that $s_2 = f(s_1)$. 
We call a GPT satisfying this property a symmetric GPT.    
\end{Def}
It is easy to see that: 
\begin{Prop}
Classical, quantum, and hyper cuboid systems are all symmetric. 
\end{Prop}
In particular, notice that, in quantum systems for any pure states $\rho_1 = \ketbra{\psi_1}{\psi_1}, \rho_2 = \ketbra{\psi_2}{\psi_2}$, there exists a unitary operator $U$ such that $\rho_2 = U\rho_1 U^\dagger$. 

We show that $S_1$ vanishes for any pure states in a symmetric GPT. 
To see this, we first show:
\begin{Lem}\label{lem:S1(s)=0}
In any GPT, there exists a pure state $s$ such that 
$S_1(s) = 0$. 
\end{Lem}
{\bf Proof} \ 
Let $e_1$ be an indecomposable and pure effect (see Proposition \ref{prop:InPureEff}), and let $u-e_1 = e_2 + \ldots + e_m$ be an indecomposable decomposition of $u-e_1$ (see Proposition \ref{lem:EdecInd}). 
Then, $M = (e_j)_{i=1}^m$ is an indecomposable measurement. 
From Proposition \ref{prop:PEzeros}, there exists a pure state $s$ such that 
$e_1(s) = 1$. 
Thus, we have $H(e_j(s)) = 0$, and $S_1(s) = 0$. 
\hfill $\blacksquare$

\begin{Prop}\label{prop:SGPTS1}
Let $\SA$ be the state space of a symmetric GPT. 
Then, $S_1(s)= 0$ for any pure state $s$. 
\end{Prop}
{\bf Proof} \ 
From Lemma \ref{lem:S1(s)=0}, there exists a pure state $s_0$ such that 
$S_1(s_0) = 0$. 
For any pure state $s$, there exists a bijective affine $f$ such that 
$s_0 = f(s)$. 
Let $M = (m_j)_j$ be an indecomposable measurement such that 
$H(m_j(s_0)) = 0$. 
Then, it is easy to see that $\tilde{M} := (\tilde{m}_j)_j$ where $\tilde{m}_j:= m_j \circ f$ is an indecomposable measurement. 
Therefore, it follows that 
$H(\tilde{m}_j(s)) = H(m_j(s_0)) = 0$.
Thus, we have proved that $S_1(s) = 0 $ for any pure state $s$.  
\hfill $\blacksquare$

Therefore, in a symmetric GPT, $S_1$ measures a pureness in some sense. 
However,  as the squared GPT shows, the converse of Proposition \ref{prop:SGPTS1} does not holds in general even among symmetric GPTs.  

\section{Concluding Remarks} 
We have discussed some distinguishability measures (especially, Kolmogorov distance and fidelity) in any GPT. 
In a similar way of quantum information theory, it will be convenient to use these measures in constructing an information theory in GPT. 
Indeed, we have reformulated no-cloning theorem and information-disturbance theorem using fidelity. 

We have also proposed and investigated three quantities related to entropies in any GPT. 
All of them are generalizations of Shannon and von Neumann entropy in classical and quantum systems, respectively. 
However, they are in general distinct quantities, as the squared system gives the example. 
The concavity of $S_1$ in any GPT holds while it breaks for $S_2$ and $S_3$ in some GPT. 
$S_2$ and $S_3$ provides a measure for pureness, while $S_1$ does not. 
However, in a symmetric GPT which satisfies the principle of {\it equality of pure states}, it follows that $S_1(s) = 0$ for any pure states $s$.  
In the attempt to find principles of our world, which is described by a quantum system at least for the present, we think that symmetric GPTs are enough to consider by assuming the principle of equality for pure states.  
However, let us remark here that both classical and quantum systems satisfy stronger principle, which we call {\it strong equality for pure states} or {\it equality for distinguishable pure states} which can be formulated as follows:
\begin{Def}(Strong equality for pure states)
We say that GPT satisfies the principle of strong equality for pure states if it satisfies the following: Let $\{s_i \in \SA_{pure}\}_{i=1}^{n}$ and $\{t_i \in \SA_{pure} \}_{i=1}^m$ (let $n \ge m$) be two distinguishable sets of pure states, i.e., there exists a measurement $M=(m_i)_i$ ($N=(n_i)_i$) such that $m_i(s_j) = \delta_{ij}$ ($n_i(t_j) = \delta_{ij}$).  
Then, there exists a bijective affine map $f$ on $\SA$ such that $t_i = f(s_i) \ (i=1,\ldots,m)$. 
\end{Def}
Notice that the squared GPT is symmetric but does not satisfies this strong equality for pure states. (For instance, consider $\{(0,0),(0,1)\}$ and $\{(0,0),(1,1)\}$.)  
It might be interesting to consider these kind of stronger conditions which classical and quantum systems satisfy. 
In particular, we don't know any principles which makes the converse of  Proposition \ref{prop:SGPTS1} to hold.

{\bf Acknowledgment} 
We would like to thank useful comments and discussions with Dr. Imafuku and Dr. Miyadera. 
Part of this work is supported by Grant-in-Aid for Young Scientists (B), The Ministry of Education, Culture, Sports, Science and Technology (MEXT) (No.20700017). 

{\it Note added}. 
Related but independent works for {\it entropies} in GPT has appeared recently in Ref.~\cite{ref:SW,ref:Ent} while completing this paper. 
We will continuously see the relations between our work and the results there in near future.   

\appendix 

\section{Proofs of some propositions}\label{app:proofs}

\noindent {\bf [Proof of Proposition \ref{prop:monD}]} \ 
Notice that for any measurement $M=\{m_i \} \in \M({\cal S}^\prime)$, and any affine map $\Lambda \in \A({\cal S},{\cal S}^\prime)$, we have another measurement $N=\{ n_i \} \in \M({\cal S})$ where $n_i := m_i \circ \Lambda $. 
Let $M=\{m_i \} \in \M({\cal S}^\prime)$ be an optimal measurement which attains the maximum: 
$$
D(\Lambda(s_1),\Lambda(s_2)) = \frac{1}{2}\sum_i |m_i(\Lambda(s_1)) - m_i(\Lambda(s_2))|.
$$
Then, we have 
$$
D(\Lambda(s_1),\Lambda(s_2)) = \frac{1}{2}\sum_i |n_i(s_1) - n_i(s_2)| \le D(s_1,s_2).  
$$  
\hfill{$\blacksquare$}

\noindent {\bf [Proof of Proposition \ref{prop:sc}] } Let $M=\{m_i\} \in \M$ be a measurement which satisfies
$$
D\Bigl( \sum_i p_i s_i, \sum_i q_i t_i \Bigr)  = \frac{1}{2} \sum_i |m_i(\sum_j p_j s_j) - m_i(\sum_j q_j t_j)|. 
$$
Then, we have 
$$
D\Bigl( \sum_i p_i s_i, \sum_i q_i t_i \Bigr)  = \frac{1}{2} \sum_i |\sum_j p_j m_i(s_j) - \sum_j q_j m_i(t_j)|
$$
\begin{eqnarray*}
&\le&  \frac{1}{2} \sum_{i,j} p_j |(m_i(s_j) -m_i(t_j)) | + \frac{1}{2}\sum_{i,j}| (p_j-q_j)| m_i(t_j)  \nonumber \\
&\le&  \sum_j p_j D(s,t) + D_c(p_i,q_i), 
\end{eqnarray*}
where we have used (i) affinity of $m_i$, (ii) triangle inequality of $| \cdot |$, and (iii) $\sum_i m_i = u$. \hfill $\blacksquare$

\noindent {\bf [Proof of Proposition \ref{prop:monF}]}\ 
The proof goes almost similar manner with that of Proposition \ref{prop:monD}, only noting a technical treatment of the infimum:  
For any $\epsilon > 0$, there exists an ``optimal" measurement $M=(m_j)_j$ such that 
$F(\Lambda(s_1),\Lambda(s_2)) + \epsilon \ge \sum_j \sqrt{m_j(\Lambda(s_1))m_j(\Lambda(s_2))}$ from the definition of the fidelity. 
By using a measurement $N=(n_i)_i $ where $n_i := m_i \circ \Lambda$, it follows that 
$F(\Lambda(s_1),\Lambda(s_2)) + \epsilon \ge \sum_j \sqrt{n_j(s_1)n_j(s_2)} \ge F(s_1,s_2)$. 
Since $\epsilon > 0$ is arbitrary, we obtain the monotonicity.  
\hfill $\blacksquare$

\noindent {\bf[Proof of Proposition \ref{prop:scF}] } 
For any $\epsilon > 0$, let $M=(m_i)_i$ be an ``optimal" measurement satisfying 
$$
F(\sum_i p_i s_i,\sum_i q_i t_i) + \epsilon \ge \sum_k \sqrt{m_k(\sum_i p_i s_i) m_k(\sum_j q_j t_j)}.
$$
Using the affinity of $m_i$ and the Schwarz inequality between vectors $(\sqrt{ p_i m_k( s_i)})_i$ and $(\sqrt{ q_i m_k( t_i)})_i$, one gets 
\begin{eqnarray*}
&&F(\sum_i p_i s_i,\sum_i q_i t_i) + \epsilon  \\ 
&\ge& \sum_k \sqrt{\sum_i p_i m_k( s_i)} \sqrt{\sum_j q_j m_k( t_j)} \\
&\ge& \sum_k \sum_i \sqrt{p_i m_k( s_i) q_i m_k( t_i)} \ge \sum_i \sqrt{p_iq_i} F(s_i,t_i). 
\end{eqnarray*}
Letting $\epsilon \to 0$, we obtain the strong concavity. 
\hfill $\blacksquare$

\noindent {\bf [Proof of Proposition \ref{prop:biF}] } \ 
(i) For any $\epsilon > 0$, let 
$M = (m_i)_i \in \M(\SA_A)$ and $N = (n_j)_j \in \M(\SA_B)$ be ``optimal" measurements 
such that $F(s_1,s_2) +\epsilon \ge \sum_i \sqrt{m_i(s_1)m_i(s_2)}$ and 
$F(t_1,t_2) + \epsilon \ge \sum_j \sqrt{n_i(t_1)n_j(t_2)}$, respectively. 
Then, $g_{ij} = m_i \otimes n_j \in \E(\SA_A\otimes \SA_B)$ gives a (joint) measurement $G = (g_{ij})_{ij} \in \M(\SA_A\otimes \SA_B)$, and 
$(F(s_1,s_2)+\epsilon)(F(t_1,t_2)+\epsilon) \ge \sum_{ij} \sqrt{g_{ij}(s_1\otimes t_1)g_{ij}(s_2 \otimes t_2)} \ge F(s_1\otimes t_1,s_2\otimes t_2)$.  

(ii) For any $\epsilon > 0$, let $M = (m_i)_i \in \SA_A$ be an ``optimal" measurement such that 
$F(s_A,t_A) + \epsilon \ge \sum_i \sqrt{m_i(s_A) m_i(t_A)}$. 
By noting that $g_i = m_i \otimes u_B$ gives a measurement on $\SA_A\otimes \SA_B$ and $g_i(s) = m_i(s_A), g_i(t) = m_i(t_A)$, one has $F(s_A,t_A) + \epsilon  \ge \sum_i \sqrt{g_i (s) g_i(t)} \ge F(s,t)$. 

(iii) The inequality $F(s_1,s_2) \ge F(s_1\otimes t,s_2 \otimes t) \ge \sum_i \sqrt{m_i(s_1)\otimes m_i(s_2)}$ follows from (i) and $F(t,t)=1$. 
To see the opposite inequality, let $G=(g_i)_i \in \M(\SA_A\otimes \SA_B)$ be an ``optimal" measurement such that $F(s_1\otimes t,s_2 \otimes t) + \epsilon \ge \sum_i \sqrt{g_i (s_1\otimes t) g_i(s_2 \otimes t)}$ for any $\epsilon > 0$. 
Then, since $m_i(s) := g_i(s\otimes t) \ (\forall s \in \SA_A)$ gives a measurement $M = (m_i)_i \in \SA_A $, we have $F(s_1\otimes t,s_2 \otimes t) + \epsilon  \ge \sum_i \sqrt{m_i(s_1)\otimes m_i(s_2)} \ge F(s_1,s_2)$.

\hfill{$\blacksquare$}

\noindent {\bf [Proof of Proposition \ref{prop:relGF}]} \ The proof is essentially the same as in \cite{ref:FG}. 
For any $\epsilon > 0$, let $M=(m_i)_i$ be an ``optimal" measurement which satisfies $F(s,t) + \epsilon \ge \sum_i \sqrt{p_iq_i}$ where $p_i := m_i(s), q_i:= m_i(t)$. 
It follows that $\sum_i (\sqrt{p_i} - \sqrt{q_i})^2 = \sum_i p_i + \sum_i q_i - 2 \sum_i \sqrt{p_i q_i} \ge 2(1-F(s,t) - \epsilon)$. 
Noting that $|\sqrt{p_i} - \sqrt{q_i}| \le |\sqrt{p_i} + \sqrt{q_i}|$, we have $
\sum_i (\sqrt{p_i} - \sqrt{q_i})^2 = \sum_i |\sqrt{p_i} - \sqrt{q_i}| |\sqrt{p_i} - \sqrt{q_i}| \le \sum |p_i - q_i| \le 2 D(s,t)$. 
Next, let $N=\{f_i\} \in \M$ be an optimal measurement which satisfies $D(s,t) = \frac{1}{2} \sum_i |r_i - s_i|$,  where $r_i = n_i(s), s_i = n_i(t)$. 
Then, we have $D(s,t)^2 = \frac{1}{4} (\sum_i |r_i - s_i|)^2 = \frac{1}{4} (\sum_i |\sqrt{r_i} - \sqrt{s_i}| |\sqrt{r_i} + \sqrt{s_i}|)^2 \le  \frac{1}{4} (\sum_i |\sqrt{r_i} - \sqrt{s_i}|^2) (\sum_i |\sqrt{r_i} + \sqrt{s_i}|^2) = \frac{1}{4} (\sum_i (\sqrt{r_i} - \sqrt{s_i})^2) (\sum_i (\sqrt{r_i} + \sqrt{s_i})^2) = \frac{1}{4} \Bigl(\sum_i r_i + \sum_i s_i - 2\sum_i \sqrt{r_is_i}\Bigr)  \Bigl(\sum_i r_i + \sum_i s_i + 2\sum_i \sqrt{r_is_i}\Bigr)  =  \Bigl(1 - \sum_i \sqrt{r_is_i}\Bigr)  \Bigl(1 +\sum_i \sqrt{r_is_i}\Bigr) = \Bigl(1 - (\sum_i \sqrt{r_is_i})^2 \Bigr)  \le 1- F(s,t)^2$, where we have used the Schwarz inequality.  
\hfill $\blacksquare$

\noindent {\bf [Proof of Proposition \ref{lem:EdecInd}]} \ 
In the proof, we use some terminology from convex geometry.
We say that a subset $C$ of a finite dimensional Euclidean space $\R^N$ is a {\em cone} if $v \in C$ and $\lambda \geq 0$ imply $\lambda v \in C$ (hence $0 \in C$).
We say that a closed convex cone $C$ is {\em pointed} if $C \cap -C = \{0\}$.
In the proof of Proposition \ref{lem:EdecInd}, we use the following fact for pointed cones:
\begin{Lem}
[{\cite[Theorem 3.3.15]{ref:Borwein}}]
\label{lem:pointed_cone}
A closed convex cone $C \subset \R^N$ is pointed if and only if there is a linear functional $f$ on $\R^N$ such that $C' = \{v \in C \mid f(v) = 1\}$ is compact and satisfies $C = \{\lambda v \mid v \in C', \lambda \geq 0\}$.
\end{Lem}
We proceed the proof of Proposition \ref{lem:EdecInd}.
Put $N = \dim(\SA) < \infty$ and let $\SA \subset V = \R^N$ (recall that now $\SA$ is finite dimensional).
Choose $s_1,\dots,s_{N+1} \in \SA$ such that $V$ is the affine hull of these $N+1$ points.
Then any affine functional on $\SA$ extends to a unique affine functional on $V$, therefore the set $\mathrm{Aff}_{\SA}^+$ of all nonnegative affine functionals $f$ on $\mathcal{S}$ can be embedded in $V' = \mathbb{R}^{N+1}$ where the $i$-th coordinate signifies the value at $s_i$.
Now the embedded image of $\mathrm{Aff}_{\SA}^+$ in $V'$ is a pointed closed convex cone, where the closedness follows since elements $f$ of the set $\mathrm{Aff}_{\SA}^+$ are characterized by closed relations among the values of $f$ at the points $s_i$.
Thus by Lemma \ref{lem:pointed_cone}, there exists a linear functional $g$ on $V'$ such that $C = \{e \in \mathrm{Aff}_{\SA}^+ \mid g(e) = 1\}$ is compact and satisfies $\mathrm{Aff}_{\SA}^+ = \{\lambda e \mid e \in C, \lambda \geq 0\}$.
Note that $C$ is convex by definition.

Let $0 \neq e \in \E(\SA)$.
Then we have $\lambda e \in C$ for some $\lambda > 0$ by the property of $C$.
Since $C$ is compact and convex, the Krein-Milman's Theorem implies that $\lambda e$ can be written as a finite convex combination $\lambda e = \sum_x p_x \overline{e}_x$ of extreme points $\overline{e}_x$ of $C$.
Since $\SA$ is compact, by taking a sufficiently small $\mu > 0$ it follows that $e_x = \mu \overline{e}_x \in \E(\SA)$ for every $x$.
Moreover, choose an integer $k > 0$ such that $k \lambda \mu \geq 1$.
Then we have a decomposition 
\begin{displaymath}
e = k \sum_x \frac{ p_x }{ k \lambda \mu } e_x \enspace,\enspace \frac{ p_x }{ k \lambda \mu } e_x \in \E(\SA)
\end{displaymath}
of $e$ into a finite collection of effects (note that $0 \leq p_x / (k \lambda \mu) \leq 1$).

Our remaining task is to show that each $q_x e_x$, where $q_x = p_x / (k \lambda \mu)$, is an indecomposable effect provided $q_x > 0$.
Let $q_x e_x = e' + e''$ with $e',e'' \in \E(\SA)$, $e',e'' \neq 0$.
Then we have $\overline{e}_x = (q_x \mu)^{-1} (e' + e'')$.
By the property of $C$, there exist $\nu',\nu'' > 0$ and $\overline{e}',\overline{e}'' \in C$ such that $e' = \nu' \overline{e}'$ and $e'' = \nu'' \overline{e}''$.
We have $\overline{e}_x = \eta' \overline{e}' + \eta'' \overline{e}''$, where $\eta' = (q_x \mu)^{-1} \nu' > 0$ and $\eta'' = (q_x \mu)^{-1} \nu'' > 0$.
Moreover, by the definition of $C$, we have
\begin{displaymath}
1 = g(\overline{e}_x) = \eta' g(\overline{e}') + \eta'' g(\overline{e}'') = \eta' + \eta'' \enspace.
\end{displaymath}
Since $\overline{e}_x$ is an extreme point of $C$, it follows that $\overline{e}_x = \overline{e}' = \overline{e}''$, therefore $e' = \nu' \overline{e}_x = (\nu'/(\mu q_x)) q_x e_x$.
Hence $q_x e_x$ is indecomposable as desired, concluding the proof of Proposition \ref{lem:EdecInd}.
\hfill $\blacksquare$

\noindent {\bf [Proof of Lemma \ref{lem:ietoie}]} \ 
It is easy to show $\tilde{e}$ is an effect. 
Let $\tilde{e} = e_1 + e_2$ be an effect decomposition of $\tilde{e}$.  
Then, $e = q e_1 + q e_2$ is an effect decomposition of $e$ since $q \le 1$. 
Since $e$ is indecomposable, there exists $c \in \R$ such that $q e_1 = c e$, or $e_1 = c\tilde{e}$. 
Thus, $\tilde{e}$ is indecomposable. 
\hfill $\blacksquare$

\noindent {\bf [Proof of Lemma4 \ref{lem:inep}]} \ 
Let $e = \lambda e_1 + (1-\lambda ) e_2$ be a convex decomposition of $e$ with $\lambda \in (0,1)$. 
It is easy to see that $e_1(s) = e_2(s) = 1$. 
Since $\lambda e_1, (1-\lambda)e_2 \in \E$ and $e$ is indecomposable, we have $\lambda e_1 = c e$ for some $c \in \R$. 
Applying this to $s$, we have $\lambda = c$, and thus $e_1 = e$. 
Therefore, $e$ is a pure effect. 
\hfill $\blacksquare$

\noindent {\bf [Proof of Lemma \ref{lem:Eindtakes0}]} \ 
First we show that an indecomposable $e$ is not constant on $\SA$.
Since $\SA$ has at least two states, the separation property of states implies that a non-constant effect $f \in \E(\SA)$ exists.
If $e$ takes constantly $c \in (0,1]$, then the decomposition $e = c f + c (u - f)$ contradicts that $e$ is indecomposable.
Hence $e$ is not constant.
Second, if $e$ does not take $0$ at any state, then we have $e(s) \geq c$ for some $c > 0$ and all $s \in \SA$ since $e$ is continuous and $\SA$ is compact.
Now the decomposition $e = c u + (e - c u)$ contradicts that $e$ is indecomposable.
Hence $e$ takes $0$ at some state.
\hfill $\blacksquare$

\noindent {\bf [Proof of Proposition \ref{prop:entropy_sq}]} \ 
First we compute $S_1(s)$ for $s = (c_1,c_2) \in \SA_{\mathrm{sq}}$.
Let $M = (m_i)_i$ be an indecomposable measurement.
To compute $S_1$, it suffices to consider the case that $M$ contains at most one effect $m_i$ of each of the four types listed in Table \ref{tab:effects_square_space}; indeed, if $m_{i_1}$ and $m_{i_2}$ are of the same type (i.e., $m_{i_2} = \lambda m_{i_1}$ for some $\lambda \in \R$), then by replacing the pair of $m_{i_1}$ and $m_{i_2}$ with $m_{i_1} + m_{i_2}$ the value of $H(m_i(s))$ is not increased.
Thus we may assume without loss of generality that $M$ consists of the four effects in Table \ref{tab:effects_square_space} with parameters $\alpha_1 = \alpha_2 = \alpha$, $\alpha_3 = \alpha_4 = \beta := 1 - \alpha$ for some $\alpha \in [0,1]$.
Now, by putting $g(x) = -x \log x$ we have
\begin{displaymath}
\begin{split}
&H(m_i(s)) \\
&= g(\alpha c_1) + g(\alpha(1-c_1)) + g(\beta c_2) + g(\beta (1-c_2)) \\
&= g(\alpha) + \alpha h(c_1) + g(\beta) + \beta h(c_2) \\
&= h(\alpha) + \alpha h(c_1) + (1 - \alpha) h(c_2) \enspace.
\end{split}
\end{displaymath}
Since the right-hand side is concave on $\alpha \in [0,1]$, it takes the minimum at either $\alpha = 0$ or $\alpha = 1$, hence we have $S_1(s) = \min[h(c_1),h(c_2)]$ as desired.

Second, we compute $S_2(s)$ for $s = (c_1,c_2) \in \SA_{\mathrm{sq}}$.
Let $\{p_x,s_x\}_x \in \PA(s)$ with $s_x = (c_{x,1},c_{x,2})$ and $M = (m_j)_j \in \M_{\mathrm{ind}}$.
Again, it suffices to consider the case that $M$ contains at most one effect $m_j$ of each of the four types listed in Table \ref{tab:effects_square_space}; indeed, if $m_{j_1}$ and $m_{j_2}$ are of the same type (in the above sense), then by replacing the pair of $m_{j_1}$ and $m_{j_2}$ with $m_{j_1} + m_{j_2}$ the value of $H(X:J)$ is not changed.
Thus we may assume without loss of generality that $M$ consists of the four effects in Table \ref{tab:effects_square_space} with parameters $\alpha_1 = \alpha_2 = \alpha$, $\alpha_3 = \alpha_4 = \beta := 1 - \alpha$ for some $\alpha \in [0,1]$.
Now a direct calculation implies that
\begin{displaymath}
\begin{split}
&H(X:J) \\
&= h(\alpha) + \alpha h(c_1) + \beta h(c_2) \\
&\quad - \sum_{x \in X} p_x (h(\alpha) + \alpha h(c_{x,1}) + \beta h(c_{x,2})) \\
&= \alpha(h(c_1) - \sum_{x} p_x h(c_{x,1})) + \beta(h(c_2) - \sum_{x} p_x h(c_{x,2})) \enspace.
\end{split}
\end{displaymath}
Since all the pure states $(c_{x,1},c_{x,2})$ in $\SA_{\mathrm{sq}}$ satisfy that $c_{x,1} \in \{0,1\}$ and $c_{x,2} \in \{0,1\}$, we have $H(X:J) = \alpha h(c_1) + \beta h(c_2) = \alpha h(c_1) + (1 - \alpha) h(c_2)$, which is independent of the given decomposition $\{p_x,s_x\}_x$ of $s$.
This implies that $S_2(s) = \sup_{X,J} H(X:J) = \max[h(c_1),h(c_2)]$, as desired.

Finally, we compute $S_3(s)$ for $s = (c_1,c_2) \in \SA_{\mathrm{sq}}$.
By the reason similar to the case of $S_1$, to compute $S_3(s)$ it suffices to consider a decomposition $\{p_x,s_x\}_{x \in X} \in \PA(s)$ such that all $s_x$ are different pure states.
Thus we may assume that $X = \{00,01,10,11\}$, $s_{00} = (0,0)$, $s_{01} = (0,1)$, $s_{10} = (1,0)$ and $s_{11} = (1,1)$.
Now by putting $p_{11} = p$ we have
\begin{displaymath}
p_{10} = c_1 - p,\, p_{01} = c_2 - p,\, p_{00} = 1 - c_1 - c_2 + p \enspace.
\end{displaymath}
In the above expression, we have $p_x \in [0,1]$ for every $x$ if and only if $p_{\mathrm{m}} \leq p \leq p_{\mathrm{M}}$, where
\begin{displaymath}
p_{\mathrm{m}} = \max[0,c_1 + c_2 - 1],\, p_{\mathrm{M}} = \min[c_1,c_2] \enspace.
\end{displaymath}
Hence we have $S_3(s) = \inf_{p_{\mathrm{m}} \leq p \leq p_{\mathrm{M}}} H(p_x)$.
Now a direct calculation shows that
\begin{displaymath}
\left( \frac{ d }{ dp } \right)^2 H(p_x) = - \sum_{x \in X} \frac{ 1 }{ p_x } < 0
\end{displaymath}
for any $p \in (p_{\mathrm{m}},p_{\mathrm{M}})$, therefore $H(p_x)$ takes the minimum at either $p = p_{\mathrm{m}}$ or $p = p_{\mathrm{M}}$: $S_3(s) = \min[H(p_x)|_{p = p_{\mathrm{m}}},H(p_x)|_{p = p_{\mathrm{M}}}]$.

First we consider the case that $c_1 \leq c_2$ and $c_1 + c_2 \leq 1$ (i.e., $s \in R_{2U}$ or $s \in R_{2B}$), therefore $p_{\mathrm{m}} = 0$ and $p_{\mathrm{M}} = c_1$.
If $p = p_{\mathrm{m}}$ then we have $(p_x)_x = (1 - c_1 - c_2, c_2, c_1, 0)$, while if $p = p_{\mathrm{M}}$ then we have $(p_x)_x = (1 - c_2, c_2 - c_1, 0, c_1)$.
This implies that
\begin{displaymath}
\begin{split}
\mathrm{diff}_{\mathrm{M-m}} &:= H(p_x)|_{p = p_{\mathrm{M}}} - H(p_x)|_{p = p_{\mathrm{m}}} \\
&= g(c_2 - c_1) + g(1 - c_2) - g(c_2) - g(1 - c_1 - c_2)
\end{split}
\end{displaymath}
where $g(a) = -a \log a$, therefore
\begin{displaymath}
\frac{\partial}{\partial c_2} \mathrm{diff}_{\mathrm{M-m}}
= \log \frac{ (1 - c_2) c_2 }{ (c_2 - c_1)(1 - c_1 - c_2) }
\end{displaymath}
which is now non-negative by the conditions for $c_1$ and $c_2$.
Since $\mathrm{diff}_{\mathrm{M-m}} = 0$ when $c_2 = 1/2$, it follows that $\mathrm{diff}_{\mathrm{M-m}} \leq 0$ and $S_3(s) = H(p_x)|_{p = p_{\mathrm{M}}}$ when $0 \leq c_2 \leq 1/2$ (i.e., $s \in R_{2B}$), and $\mathrm{diff}_{\mathrm{M-m}} \geq 0$ and $S_3(s) = H(p_x)|_{p = p_{\mathrm{m}}}$ when $1/2 \leq c_2 \leq 1$ (i.e., $s \in R_{2U}$).
Hence the expressions of $S_3(s)$ in (\ref{eq:S3sq}) for $s \in R_{2U}$ and $s \in R_{2B}$ are proved.
The claim for the remaining cases follow by considering suitable symmetry of the state space $\SA_{\mathrm{sq}}$.
\hfill $\blacksquare$

\noindent {\bf [Proof of Proposition \ref{prop:relationofS}]} \ 
The first inequality $S_1(s) \leq S_2(s)$ is obvious by (\ref{eq:S1sq}) and (\ref{eq:S2sq}).
For the second inequality $S_2(s) \leq S_3(s)$, by symmetry, we may assume without loss of generality that $s = (c_1,c_2) \in R_{2U}$, i.e., $1/2 \leq c_2 \leq 1 - c_1$.
This condition implies that $h(c_1) \leq h(c_2)$, therefore $S_2(s) = h(c_2)$.
On the other hand, (\ref{eq:S3sq}) implies that $S_3(s) = g(c_1) + g(c_2) + g(1 - c_1 - c_2)$, where $g(x) = -x \log x$.
Thus we have
\begin{displaymath}
S_3(s) - S_2(s) = g(c_1) + g(1 - c_1 - c_2) - g(1 - c_2)
\end{displaymath}
which is a decreasing function of $c_2$ in this range, while $S_3(s) - S_2(s) = 0$ when $c_2 = 1 - c_1$.
This implies that $S_3(s) \geq S_2(s)$ for any $s \in R_{2U}$, hence the claim holds.
\hfill $\blacksquare$

\noindent {\bf [Proof of Lemma \ref{lem:lemma_weak_concavity}]} \ 
For each $x \in X$, let $\{q^x_y,t^x_y\}_{y \in Y_x} \in \D(s_x)$ and $M_x = (m^x_j)_{j \in J_x} \in \M$.
Then we have $\{p_x q^x_y, t^x_y\}_{x \in X,y \in Y_x} \in \D(s)$ and $M' = (\pi_x m^x_j)_{x \in X,j \in J_x} \in \mathcal{M}$.
Let $Z = \{(x,y) \mid x \in X, y \in Y_x\}$ and $K = \{(x,j) \mid x \in X, j \in J_x\}$ denote the index sets of these ensembles, respectively.
Then, by putting $h(a) = - a \log a$ we have
\begin{displaymath}
\begin{split}
&H(Z:K)
= H(K) - H(K \mid Z) \\
&= \sum_{(x,j) \in K} h(\pi_x m^x_j(s)) - \sum_{\substack{(x',y) \in Z \\ (x,j) \in K}} p_{x'} q^{x'}_y h(\pi_x m^x_j(t^{x'}_y)) \\
&= \sum_{(x,j)} m^x_j(s) h(\pi_x) - \sum_{(x',y),(x,j)} p_{x'} q^{x'}_y m^x_j(t^{x'}_y) h(\pi_x) \\
&\quad + \sum_{(x,j)} \pi_x h(m^x_j(s)) - \sum_{(x',y),(x,j)} p_{x'} q^{x'}_y \pi_x h(m^x_j(t^{x'}_y)) \\
&= (1 - \sum_{(x',y)} p_{x'} q^{x'}_y) \sum_{x} h(\pi_x) \\
&\quad + \sum_{(x,j)} \pi_x \left( h(m^x_j(s)) - \sum_{(x',y)} p_{x'} q^{x'}_y h(m^x_j(t^{x'}_y)) \right) \\
&= \sum_{(x,j)} \pi_x \left( h(m^x_j(s)) - \sum_{(x',y)} p_{x'} q^{x'}_y h(m^x_j(t^{x'}_y)) \right) \enspace.
\end{split}
\end{displaymath}
Since $m^x_j(s) = p_x m^x_j(s_x) + \sum_{(x',y) \in Z;\,x' \neq x} p_{x'} q^{x'}_y m^x_j(t^{x'}_y)$ and $h(a)$ is concave on $a \in (0,1)$, we have $h(m^x_j(s)) \geq p_x h(m^x_j(s_x)) + \sum_{(x',y);\,x' \neq x} p_{x'} q^{x'}_y h(m^x_j(t^{x'}_y))$, therefore
\begin{displaymath}
\begin{split}
H(Z:K)
&\geq \sum_{(x,j)} \pi_x \left( p_x h(m^x_j(s_x)) - \sum_{y \in Y_x} p_x q^x_y h(m^x_j(t^x_y)) \right) \\
&= \sum_{x \in X} \pi_x p_x H(Y_x:J_x) \enspace.
\end{split}
\end{displaymath}
By taking the supremum over all $\{q^x_y,t^x_y\}_{y \in Y_x}$ and $M_x$, $x \in X$ (see Lemma \ref{lem:S3}), it follows that
\begin{displaymath}
S_2(s) \geq \sup H(Z:K)
\geq \sum_{x \in X} \pi_x p_x S_2(s_x) \enspace.
\end{displaymath}
Hence Lemma \ref{lem:lemma_weak_concavity} holds.
\hfill $\blacksquare$

\noindent {\bf [Proof of Lemma \ref{lem:S1andpseudoclassical_1}]} \ 
We use induction on the number $N$ of indices $j$ such that $x_j \not\in \{0,1/k\}$.
The claim is trivial if $N = 0$, while it cannot happen that $N = 1$.
We assume $N \geq 2$, and $0 < x_1 \leq x_2 < 1/k$ by symmetry.
Now if $x_1 + x_2 \leq 1/k$, then we have $h(x_1) + h(x_2) \geq h(x_1 + x_2)$, therefore $H(x) \geq H(y)$ where $y = (x_1 + x_2,x_3,\dots,x_{\ell})$.
On the other hand, if $x_1 + x_2 > 1/k$, then we have $h(x_1) + h(x_2) \geq h(x_1 + x_2 - 1/k) + h(1/k)$, therefore $H(x) \geq H(y)$ where $y = (x_1 + x_2 - 1/k,1/k,x_3,\dots,x_{\ell})$.
In any case, we have $H(y) \geq \log k$ by the induction hypothesis.
Hence $H(x) \geq \log k$ as desired. \hfill $\blacksquare$

\noindent {\bf [Proof of Lemma \ref{lem:S1andpseudoclassical_2}]} \ 
Choose $t_1,t_2,\dots,t_n \in \mathcal{S}$ ($n = \dim \mathcal{S} + 1$) such that these are affine independent.
Then any element $t$ of $\mathcal{S}$ has a unique expression $t = \lambda_1 t_1 + \cdots + \lambda_n t_n$, $\sum_j \lambda_j = 1$.
Let $\varphi:\mathcal{S} \to \mathbb{R}^n$ denote the map $t \mapsto (\lambda_1,\dots,\lambda_n)$.
Since $\mathcal{S}$ is a topological subspace of a finite-dimensional Euclidean space, $\mathcal{S}$ and $\varphi(\mathcal{S}) \subset \mathbb{R}^n$ are homeomorphic via $\varphi$.
By identifying $\mathcal{S}$ with $\varphi(\mathcal{S})$ in this way, the map $f_s$ is written as $f_s(e;\lambda_1,\dots,\lambda_n) = \lambda_1 e(t_1) + \cdots + \lambda_n e(t_n)$.
This implies that $f_s$ is continuous, since both $(e;\lambda_1,\dots,\lambda_n) \mapsto e(t_j)$ and $(e;\lambda_1,\dots,\lambda_n) \mapsto \lambda_j$ are continuous. \hfill $\blacksquare$

\section{GPT without complete measurements. }\label{app:AppB} 
Here we give an example of a GPT that has no complete measurements.
First note that any nonzero extreme effect $e$ takes $1$ at some state, as otherwise we have a nontrivial expression $e = c (c^{-1} e) + (1-c)0$ as a convex combination of effects, where $c = \sup_{s \in \SA} e(s) \in (0,1)$.

We consider a GPT with state space $\SA$ which is the convex hull of four points $(0,0)$, $(1,0)$, $(0,1)$, $(2,2)$ in $\R^2$.
Then by the above observation and Lemma \ref{lem:Eindtakes0}, each indecomposable extreme effect $e$ takes $0$ at an edge of $\SA$ and takes $1$ at some state (precisely, at the vertex of $\SA$ farthest from the edge).
Thus there are four indecomposable extreme effects in total, as listed in Table \ref{tab:effects_skew_square_space}.
Now it is obvious that no complete measurements exist in the GPT, since the sum of the values of indecomposable extreme effects at the state $(0,0)$ cannot equal to $1$.
\begin{table}[htb]
\centering
\caption{Indecomposable extreme effects in Appendix \ref{app:AppB}}
\label{tab:effects_skew_square_space}
\begin{tabular}{|c||c|c|c|c|} \hline
& \multicolumn{4}{|c|}{value at} \\
effect & $(0,0)$ & $(1,0)$ & $(0,1)$ & $(2,2)$ \\ \hline\hline
$e_1$ & $0$ & $0$ & $1/2$ & $1$ \\ \hline
$e_2$ & $0$ & $1/2$ & $0$ & $1$ \\ \hline
$e_3$ & $2/3$ & $0$ & $1$ & $0$ \\ \hline
$e_4$ & $2/3$ & $1$ & $0$ & $0$ \\ \hline
\end{tabular}
\end{table}


\end{document}